\documentclass[lettersize,journal]{IEEEtran}
\usepackage{amsmath,amsfonts}
\usepackage{algorithmic}
\usepackage{array}
\usepackage[caption=false,font=normalsize,labelfont=sf,textfont=sf]{subfig}
\usepackage{textcomp}
\usepackage{stfloats}
\usepackage{url}
\usepackage{verbatim}
\usepackage{graphicx}
\def\BibTeX{{\rm B\kern-.05em{\sc i\kern-.025em b}\kern-.08em
    T\kern-.1667em\lower.7ex\hbox{E}\kern-.125emX}}
\usepackage{balance}

\usepackage[switch]{lineno}
\usepackage{amssymb}
\usepackage{multirow}
\usepackage{booktabs}

\usepackage{algorithm}
\usepackage{xcolor}
\usepackage{xspace}
\usepackage{flushend}
\newcommand{\ie}{\emph{i.e.,}\xspace}

\newcommand{\eg}{\emph{e.g.,}\xspace}

\newcommand{\fig}{Fig.\xspace}
\newcommand{\baby}{{SIDSBR}\xspace}

\begin{document}
\title{A Survey on Side Information-driven Session-based Recommendation: From a Data-centric Perspective}

\author{
Xiaokun Zhang, Bo Xu, Chenliang Li, Bowei He, Hongfei Lin, Chen Ma, and Fenglong Ma
        % John~Doe,~\IEEEmembership{Fellow,~OSA,}
        % and~Jane~Doe,~\IEEEmembership{Life~Fellow,~IEEE}% <-this % stops a space
\IEEEcompsocitemizethanks{
\IEEEcompsocthanksitem Manuscript received 10 Oct. 2024; revised 6 Mar. 2025; accepted 28 Apr. 2025. This work was supported by the Natural Science Foundation of China (No.62376051, No.62076046), the Early Career Scheme (No. CityU 21219323) and the General Research Fund (No. CityU 11220324) of the University Grants Committee (UGC), and the NSFC Young Scientists Fund (No. 9240127). (Corresponding authors: Bo Xu and Chen Ma.)
\IEEEcompsocthanksitem Xiaokun Zhang, Bo Xu and Hongfei Lin are with the School of Computer Science and Technology, Dalian University of Technology, Dalian, China.
E-mail: dawnkun1993@gmail.com, xubo@dlut.edu.cn, hflin@dlut.edu.cn
% \{xubo, hflin\}@dlut.edu.cn
\IEEEcompsocthanksitem Bowei He, Chen Ma are with the Department of Computer Science, City University of Hong Kong.
E-mail: boweihe2-c@my.cityu.edu.hk, chenma@cityu.edu.hk
% \IEEEcompsocthanksitem Bo Xu and Hongfei Lin are with the School of Computer Science and Technology, Dalian University of Technology, Dalian, China.
% \protect\\
% note need leading \protect in front of \\ to get a newline within \thanks as
% \\ is fragile and will error, could use \hfil\break instead.
% E-mail: \{xubo, hflin\}@dlut.edu.cn
\IEEEcompsocthanksitem Chenliang Li is with the School of Cyber Science and Engineering, Wuhan University, Wuhan, China. E-mail: cllee@whu.edu.cn
\IEEEcompsocthanksitem Fenglong Ma is with the College of Information Sciences and Technology, Pennsylvania State University, Pennsylvania, USA. E-mail: fenglong@psu.edu
% \IEEEcompsocthanksitem Digital Object Identifier 10.1109/TKDE.2023.3309995
}% <-this % stops an unwanted space

}

\markboth{IEEE TRANSACTIONS ON KNOWLEDGE AND DATA ENGINEERING}
{ZHANG \MakeLowercase{\textit{et al.}}: }

\maketitle

\begin{abstract}
Session-based recommendation is gaining increasing attention due to its practical value in predicting the intents of anonymous users based on limited behaviors. Emerging efforts incorporate various side information to alleviate inherent data scarcity issues in this task, leading to impressive performance improvements. The core of side information-driven session-based recommendation is the discovery and utilization of diverse data. In this survey, we provide a comprehensive review of this task from a data-centric perspective. Specifically, this survey commences with a clear formulation of the task. This is followed by a detailed exploration of various benchmarks rich in side information that are pivotal for advancing research in this field. Afterwards, we delve into how different types of side information enhance the task, underscoring data characteristics and utility. Moreover, we discuss the usage of various side information, including data encoding, data injection, and involved techniques. A systematic review of research progress is then presented, with the taxonomy by the types of side information. Finally, we summarize the current limitations and present the future prospects of this vibrant topic.
\end{abstract}

\begin{IEEEkeywords}
Session-based recommendation, Side information, Benchmarks, Data-centric perspective.
\end{IEEEkeywords}

\section{Introduction}

% Recommender system (RS), as an indispensable tool to combat information overload, plays a vital role in the current information era. Assuming access to user identity information, conventional RSs rely on user profiles and long-term behaviors to predict their preferences. In today's information age, unfortunately, obtaining user profiles is becoming increasingly challenging due to increasingly stringent privacy policies. 
% % In many instances, RSs can only access short behavior sequences of anonymous users, commonly referred to as sessions.  
% To address this prevalent scenario, session-based recommendation (SBR) is proposed to predict next items of interest for \textbf{anonymous} users within \textbf{short} sessions~\cite{GRU4Rec,NARM}. Owing its significant application value, SBR has garnered widespread attention since its inception~\cite{Wang@CS2022}. 

\IEEEPARstart{R}{ecommender} systems (RS), as an effective tool to handle information overload, play a vital role in the current information era. Generally, we can summarize the utility of RS from two views, \ie information-consumer view and information-provider view. On one hand, for information consumers (\ie users), RS can significantly reduce their time and effort required to find information of interest among vast amounts of data, thus lowering the cost of obtaining valuable information. On the other hand, as for information providers, these systems can effectively identify potential consumers for their products, thereby facilitating information distribution and transaction completion. 

% General recommender systems typically rely on \textit{users' profiles} and \textit{long-term behaviors} to provide personalized recommendations~\cite{Zhang@CS2019}. 
General recommender systems typically rely on \textit{users' profiles} and \textit{long-term behaviors} to mine their behavior patterns, like collaborative relations and co-occurrence signals, for personalized recommendations~\cite{Zhang@CS2019}. However, due to stringent privacy policies, it has become increasingly difficult to access such sensitive identity information, leading to the unsatisfactory performance of general recommendation methods. 

\begin{figure}[t]
  \centering
  \includegraphics[width=0.95\linewidth]{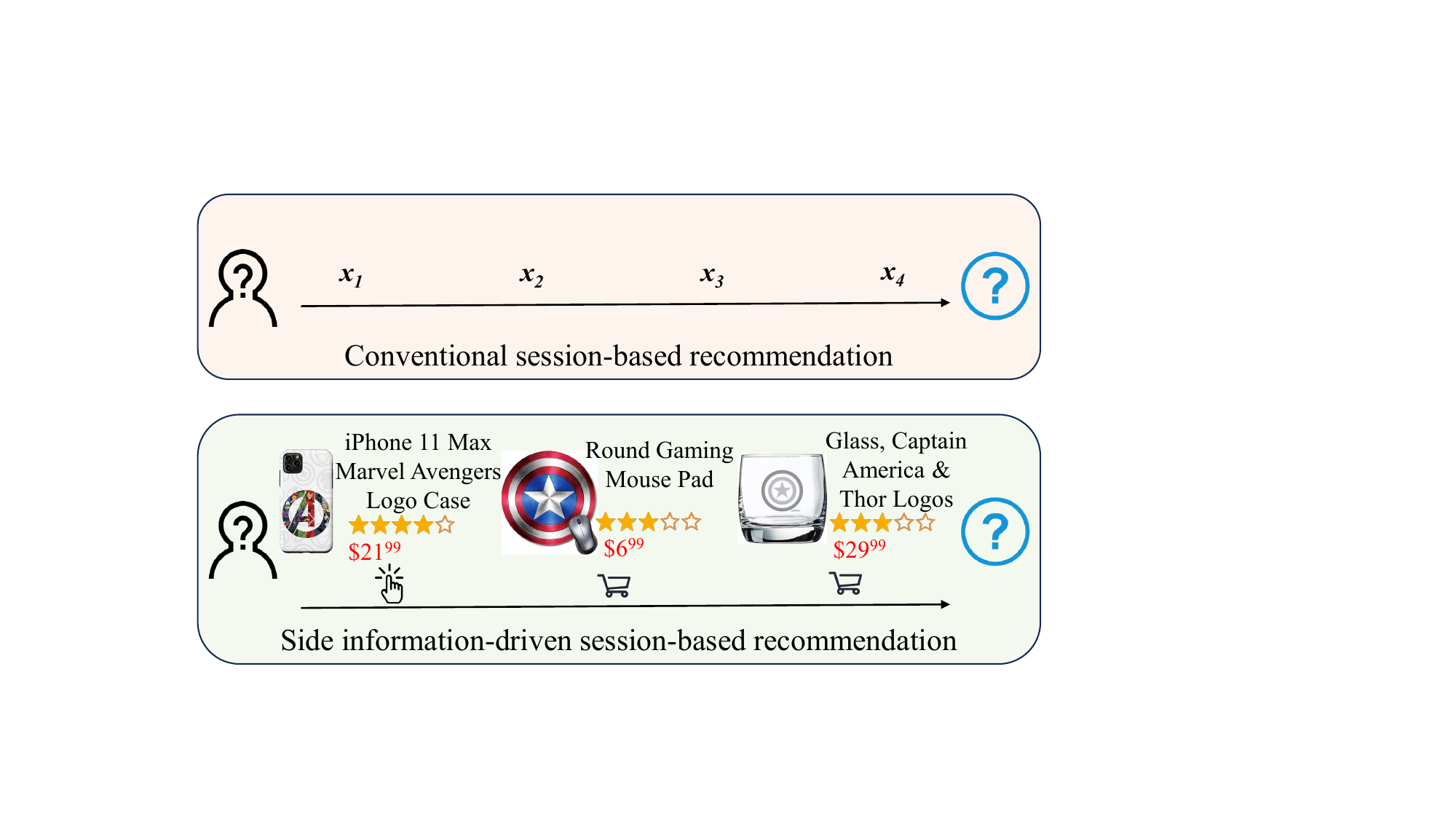}
  \caption{Conventional SBR v.s. side information-driven SBR.}\label{intro}
% \vspace{-0.1in}
\end{figure}

Session-based recommendation (SBR) is proposed to address this prevalent scenario, where SBR can predict the next items of interest for an \textit{anonymous} user based on her limited behaviors within a \textit{short} session~\cite{GRU4Rec,NARM}. Owing to possessing significantly practical value, SBR has garnered widespread attention since its inception~\cite{GRU4Rec,NARM,Wang@CS2022}. In the meantime, unfortunately, SBR inevitably suffers from serious data
scarcity issues~\cite{DHCN,Yang@SIGIR2023}. As shown in the first row of \fig~\ref{intro}, conventional SBR mainly relies on a limited number of user-item interactions merely represented by item IDs to infer a user's intents. In fact, these methods can just mine co-occurrence associations among items while failing to reveal insightful user behavior intents~\cite{MMSBR,DIMO}. As a result, this paradigm greatly limits the effectiveness of SBR.

To this end, as shown in the second row of \fig~\ref{intro}, emerging efforts tend to incorporate various types of side information, such as item images~\cite{P-RNN,MMSBR}, title text~\cite{liu@WWW2020,hou@KDD2022}, price~\cite{CoHHN,BiPNet} and behavior types~\cite{meng@SIGIR2020,Yuan@ICDE2022}, to facilitate user intent understanding under the circumstances of SBR. The incorporation of side information into SBR can undoubtedly enrich session data, and its advantages can be summarized into the following aspects:  (1) Depicting item detailed features. For example, images can convey a clothing's style, such as whether it is tight or loose. Such detailed features offer a comprehensive portrayal of an item, encompassing crucial factors in determining users' choices. (2) Revealing user fine-grained preferences. As shown in the second row of \fig~\ref{intro}, we could infer that the user may be a Marvel fan since there are commonly Marvel factors in the images or text of items this user has bought. Such insights about user fine-grained preferences open up the possibility of providing highly personalized services, like recommending a T-shirt featuring an Iron Man logo to this user.

Despite the increasing trend of incorporating side information into SBR, unfortunately, there is no corresponding survey thoroughly reviewing this vibrant field of \underline{s}ide \underline{i}nformation-\underline{d}riven \underline{s}ession-\underline{b}ased \underline{r}ecommendation abbreviated as \baby. Especially, the data is of great significance in the domain, as it affects the design and evaluation of recommendation algorithms. Under SBR scenarios, different datasets apply distinct methods to generate session data (please refer to Section \ref{benchmark} for details on session generation), which greatly impacts recommendation performance. More importantly, data determines from what view corresponding approaches aim to model user behaviors. Specifically, different types of side information portray user behaviors from different aspects. For instance, the item price signifies a user's price sensitivity, while the image converts her preferences on an item's appearance. As a result, there is an urgent need to examine the \baby with a focus on data perspective.

In this survey, therefore, we aim to provide a comprehensive review of the \baby from a data-centric perspective. Specifically, we first give a clear formulation of \baby and clarify the differences between two frequently confused tasks, \ie session-based recommendation and sequential recommendation. Available benchmarks supporting the research of \baby are then outlined, which elaborates on the types of side information they contain. Next, we delve into the data characteristics and utility of each type of information, presenting their potential to improve the SBR task. The usage of side information is then discussed, with an emphasis on data encoding, data injection, and involved techniques. We thoroughly examine current methods for \baby using a new taxonomy based on the types of information they rely on, offering convenience for future work in replicating and comparing these methods. Finally, we discuss the limitations of current \baby approaches and look forward to the future directions of this vibrant topic.
The main contributions of this survey can be summarized as follows:
\begin{itemize}
    \item We comprehensively review side-information-driven session-based recommendations from a data-centric perspective. To the best of our knowledge, this survey marks the first effort to examine this vibrant topic. 
    
    \item This survey thoroughly presents formulation, available benchmarks, data utility and usage, as well as research progress for side information-driven session-based recommendation. 

    \item We highlight and discuss several promising research directions for this topic, looking forward to inspiring the community.
    
\end{itemize}

\textit{Paper collection.}
This survey provides a comprehensive review of over 60 papers that focus on the topic of side information-driven session-based recommendation. For literature collection, DBLP\footnote{\url{https://dblp.org}} and Google Scholar\footnote{\url{https://scholar.google.com}} are primary means of searching for relevant works. Specifically, we rely on keywords, such as ``session-based recommendation'', ``session recommendation'' and ``sequential recommendation'', to search related works from these literature libraries. Besides, we focus on works published after 2016, following the first proposal and wide acceptance of SBR in GRU4Rec~\cite{GRU4Rec} presented at ICLR 2016. To ensure the quality of collected works, we prioritize papers from top-tier conferences and journals, including SIGIR, KDD, WWW, ICDE, AAAI, IJCAI, WSDM, CIKM, ICDM, RecSys, IEEE TKDE, and ACM TOIS. Additionally, we also pay attention to the unpublished preprints on arXiv, selecting those with interesting ideas or novel techniques, especially the large language models-related methods, to present a more inclusive panorama for \baby.

\textit{Differences from existing surveys.}
Our survey distinguishes itself from existing works of several key areas, including reviews on sequential recommendation (SR)~\cite{Wang@IJCAI2019, Fang@TOIS2020}, multi-modal recommendation (MMRec)~\cite{Liu@ComSur2025, Sun@ECRA2019} and session-based recommendation (SBR)~\cite{ludewig@UMUAI2020,Wang@CS2022, Li@ComSur2025}. Firstly, as clarified in Section \ref{disSBRSR}, SBR is different from SR in several perspectives, including task definition, data distribution, and experiment settings. Different from existing reviews on SR, this survey is specifically tailored to explore various approaches that are uniquely suited to the task of SBR. Secondly, MMRec methods typically require user profiles to guide model implementation, leading to their inapplicability to SBR. Moreover, different from current MMRec mainly focusing on text and images, this survey extends the scope to include extra information particularly prevalent in SBR, such as time and category. The time information sheds light on user dynamic intents within a session~\cite{Alexander@arXiv2017,Yang@ICME2020}, while category information is popular in SBR to reveal hierarchical transition patterns among items~\cite{cai@SIGIR2021,lin@WWW2022}. Thirdly, existing reviews on SBR~\cite{ludewig@UMUAI2020,Wang@CS2022} mainly focus on a single kind of information, \ie~item ID, while ignoring the significance of various side information in revealing user intents within short sessions. In contrast, our survey embraces a wider range of data types, clarifying the potential of rich data in revealing user intents. This comprehensive review on SBR from a data-centric perspective adapts to current requirements of the community, facilitating understand, exploration and implementation of researchers and practitioners on this vibrant topic. 

\textit{Content arrangement.} 
The remaining content of this survey is organized as follows. Section \ref{taskFor} presents the paradigm of side information-driven session-based recommendation and details the distinction between session-based recommendation and sequential recommendation. Section \ref{benchmark} introduces the available datasets to the task of \baby in detail. Section \ref{dataUti} outlines the characteristics and utility of various types of side information on SBR. Section \ref{Usage} presents the encoding manners of side information, illustrates the methods for incorporating side information from item-level, session-level and prompt-level, and summarizes techniques involved in the domain. Section \ref{ResearchP} elaborates on the relevant works in \baby from a new  taxonomy, \ie types of side information they used. Section \ref{openf} discusses the limitations of existing \baby and provides several promising research directions about this popular field. Section \ref{conclusion} concludes this survey. 

\section{Task Formulation}\label{taskFor}

\subsection{Paradigm of \baby}

\begin{figure*}[t]
  \centering
  \includegraphics[width=0.85\linewidth]{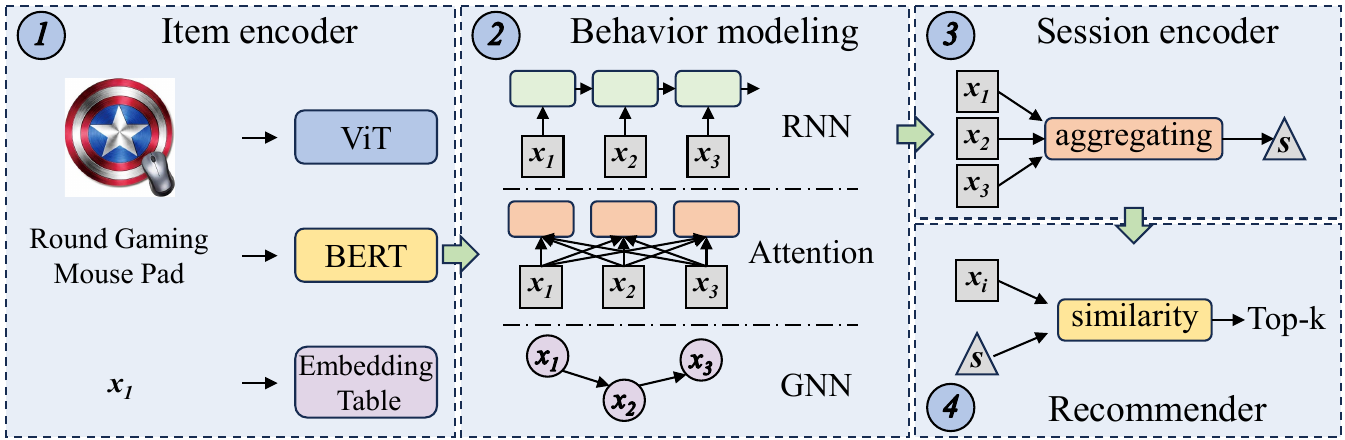}
  \caption{The workflow of side information-driven session-based recommendation.}\label{workflow}
% \vspace{-0.1in}
\end{figure*}

The emergence of session-based recommendation has coincided with the rise of deep learning techniques, leading to the dominance of neural models in this field. Generally, neural models focus on obtaining item and session embeddings by modeling user behaviors within sessions, basing recommendations on their similarity. As shown in \fig~\ref{workflow}, the common workflow of \baby can be summed up as four steps: (1) Item encoder, which converts item information into dense embeddings. For instance, look-up embedding table is usually used to map discrete item IDs into ID embeddings. (2) Behavior modeling, which captures transition patterns among items in sessions for item embedding learning. Various neural structures are modified in this stage to handle \baby, such as Recurrent Neural Network (RNN)~\cite{GRU4Rec, P-RNN, HidasiK@CIKM2018}, attention mechanism~\cite{STAMP,ATEM, SASRec, BERT4Rec} or Graph Neural Network (GNN)~\cite{SR-GNN, GCE-GNN, Star@CIKM2020}. (3) Session encoder, which encodes intents of the anonymous user as a dense vector by aggregating item embeddings within the session. (4) Recommender, which usually calculates the similarity between session embedding and item embeddings, \eg cosine similarity, to formulate the recommendation list.

Formally, let $\mathcal{I}$ denote an item set, where $n=|\mathcal{I}|$ is the total number of unique items. An item $x_i \in \mathcal{I}$ may contain various kinds of side information, such as category, text (title and description), image, and so on. 
% That is, $x_i = \{x^{id}, x^{cat}, x^{pri}, x^{txt}, x^{img}, ...... \}$, where `$id$' `$cat$', `$pri$', `$txt$', `$img$' signify item ID, category, price, text and image, respectively. 
Besides, a session ${S}$ = [$x_1, x_2, ..., x_t$] ($x_j \in \mathcal{I}$) consists of $t$ items an anonymous user interacted with within a short period. Current methods of \baby typically encode item information as dense embeddings $\textbf{v}_i \in \mathbb{R}^d$ ($d$ is the embedding dimension), model item transition relations throughout sessions, derive session embedding $\textbf{s}$ via various approaches, and evaluate similarity scores between the current session embedding and candidate item embeddings. Finally, the items with top-$k$ scores are shown to users for personalized recommendation.

\subsection{Distinction between SBR and SR}\label{disSBRSR}
Some researchers and practitioners in the community often indiscriminately conflate two popular recommendation tasks, namely session-based recommendation (SBR) and sequential recommendation (SR). As a result, it is crucial to clarify the distinction between SBR and SR before delving into the \baby in detail. Specifically, we will comprehensively distinguish between SBR and SR from a macro-view to a micro-view across three aspects: task definition, data characteristics and experimental implementations. 

For the task definition, SBR is dedicated to predicting actions of an \textit{anonymous} user based solely on her \textit{short-term} behaviors happened in a short period, such as 30 minutes or one day~\cite{Yuan@WWW2020,Wei@WWW2022,Xin@WSDM2024}. In contrast, SR assumes the availability of user profiles, and can employ rich \textit{user information} like demographics and their entire \textit{long-term behaviors} spanning from several months to several years to provide personalized suggestion~\cite{Li@SIGIR2024,FineRec,Li@TKDE2024}. 

From the perspective of data characteristics, the sequence length in SBR is much shorter than that of SR~\cite{Zhang@AAAI2024,Choi@SIGIR2024,Wang@SDM2021}. Generally, a session in SBR commonly contains no more than five items, whereas the sequence length in SR is usually longer than ten. Additionally, there are significantly more sequences in SBR compared to SR. Under the scenarios of SR, the number of sequences corresponds to the number of users. However, in SBR situations, a single user sequence may be divided into multiple sessions based on pre-defined time span, resulting in a large amount of unique sessions. These characteristics, \ie short and numerous sessions, also make SBR inherently more challenging than SR. 

Regarding experimental implementations, SR relies on user profiles, such as user ID or demographics, to derive user embeddings (\ie sequence embeddings). Conversely, SBR aggregates item embeddings within a session to obtain the session embedding without accessing any extra user information. Moreover, SR typically applies a leave-one-out operation to split all sequences into training, validation and test sets~\cite{Xie@ICDE2022,Li@TOIS2024,Lin@WWW2023}. This means that for a single sequence, the last item is used for testing, the penultimate item for validation, and the remaining for training. In contrast, SBR formulates experiment sets via grouping sessions instead of splitting any single session~\cite{MTD,Chen@CIKM2019,GC-SAN}. 

In order to cover as many methods as possible suited to the task of SBR, this survey includes SBR methods as well as certain SR models that can be modified with reasonable efforts to handle the situation of SBR. Specifically, the selected SR models should be effective when trained using the aforementioned SBR experimental implementations. This incorporation of certain SR methods facilitates the expansion of our review's scope, providing a broad reference for the community.

\section{Benchmarks}\label{benchmark}
Datasets, as collections of user-item interaction data, are the foundation for recommender system research. They serve as the essential groundwork for designing, training, and evaluating recommendation algorithms. This is particularly true for \baby, where the types of information in a dataset dictate research questions it supports. For instance, studying user price sensitivity is infeasible with datasets lacking price information. Similarly, exploring multi-modal recommendation algorithms is only possible with datasets containing multi-modal information. Unfortunately, there is no work comprehensively reviewing important datasets in the field. To this end, this section provides a detailed introduction to the available datasets for promoting future research in \baby.

\begin{table*}[th]
\small
\tabcolsep 0.09in 
  % \centering
  \caption{Available datasets for side information-driven SBR.}

\begin{tabular}{c|c|c|c|    c|c|c|c|    c|c|c|c}
\toprule
{Dataset}  & {Time$^*$} & {Category}    & {Brand}   & {Price}   & {Title}   & {Description} & {Image}    & {Address}   & {Rating$^*$}  & {Review$^*$}  & {Behavior$^*$}\\ 
\hline
{Amazon} & {$\heartsuit$} & {$\spadesuit$}  & {$\spadesuit$}  & {$\heartsuit$}  & {$\spadesuit$}  &{$\spadesuit$}   & {$\spadesuit$} & {}   & {$\heartsuit$}  &{$\spadesuit$}   & {}  \\ 
 \hline
{Beer} & {$\heartsuit$}  & {$\spadesuit$}                     & {} & {} & {$\spadesuit$}                     & {} & {} & {}   & {$\heartsuit$}  & {$\spadesuit$} & {} \\ 
\hline
{Cosmetics} &{$\heartsuit$} & {$\heartsuit$}  & {$\spadesuit$}  & {$\heartsuit$}  & {}    & {}    & {}   & {}    & {}    & {}    & {$\spadesuit$}  \\ 
\hline
{Diginetica} & {$\heartsuit$}   & {$\heartsuit$}  & {}    & {$\heartsuit$}  & {$\heartsuit$}  & {}    & {}   & {}   &{} & {}    & {}    \\
\hline
{Foursquare}       & {$\heartsuit$}    & {$\spadesuit$}                     & {} & {} & {} & {} & {}     & {$\heartsuit$}  & {}   & {}  & {}   \\ 
\hline
{Goodreads}  & {$\heartsuit$} &{} &{} &{} &{$\spadesuit$}   & {$\spadesuit$}  & {$\spadesuit$}  & {}  & {$\heartsuit$}  & {$\spadesuit$}  & {}    \\ 
\hline
{Instacart} & {$\heartsuit$}  &{$\spadesuit$} &{} &{} &{$\spadesuit$}   & {}  & {}  & {}  & {}  & {}  & {$\spadesuit$}    \\ 
\hline
{JingDong} & {$\heartsuit$} & {$\heartsuit$}                     & {} & {} & {} & {} & {}     & {}  & {} & {} & {$\spadesuit$}   \\ 
\hline
% {KKBOX} & {$\heartsuit$} & {$\heartsuit$}                     & {} & {} & {} & {} & {}       & {} & {} & {$\spadesuit$}   \\ 
% \hline
{LastFM}    & {$\heartsuit$}   & {$\spadesuit$}   & {} & {} & {$\spadesuit$}    & {}    & {}  & {}   & {}  & {}  & {}    \\ 
\hline
{Movielens}   & {$\heartsuit$}   & {$\spadesuit$}   & {} & {} & {$\spadesuit$}    & {}    & {}   & {}   & {$\heartsuit$}  & {}  & {}    \\ 
\hline
{NineRec} & {$\heartsuit$}  &                     & {} & {} & {$\spadesuit$}                     & {$\spadesuit$} & {$\spadesuit$}  & {}  & {}  & {} & {} \\ 
\hline
{RetailRocket} & {$\heartsuit$}  & {$\heartsuit$}   & {} & {} & {} & {} & {}     & {}  & {} & {} & {$\spadesuit$}  \\ 
\hline
{Steam}  & {$\heartsuit$}  & {$\spadesuit$}  & {}    & {$\heartsuit$}  & {$\spadesuit$}  & {} & {}  & {}  & {} & {$\spadesuit$} & {}    \\ 
\hline
{Taobao}     & {$\heartsuit$}    & {$\heartsuit$}                     & {} & {} & {} & {} & {}      & {} & {} & {} & {$\spadesuit$}                 \\ 
\hline
{Tmall}       & {$\heartsuit$}     & {$\heartsuit$}                     & {$\heartsuit$}                     & {} & {} & {} & {}   & {}  & {} & {} & {$\spadesuit$}  \\ 
\hline
{Trivago}   & {$\heartsuit$}     & {} & {} & {$\heartsuit$}                     & {} & {} & {}     & {}  & {} & {} & {$\spadesuit$} \\ 
\hline
{XING} & {$\heartsuit$}  &{$\spadesuit$} &{} &{} &{$\spadesuit$}   & {}  & {}   & {$\heartsuit$} & {}  & {}  & {$\spadesuit$}    \\ 
\hline
{Yelp}       & {$\heartsuit$}    & {$\spadesuit$}                     & {} & {} & {} & {} & {}     & {$\heartsuit$}  & {$\heartsuit$}   & {$\spadesuit$}  & {}   \\ 
\hline
{Yoochoose}   & {$\heartsuit$}    & {$\heartsuit$}                     & {} & {$\heartsuit$}                     & {} & {} & {}   & {}   & {} & {} & {}  \\
% \midrule
% \multicolumn{11}{p{\linewidth}}{Amazon contains many sub-datasets covering various domains such as Beauty, Cellphones and Accessories, Toys and Games and so on. Similarly, JingDong also contains two sub-datasets Applicances and Computers}          \\ 
\bottomrule
\end{tabular}\vspace{3px}
{The symbol $^*$ denotes \textbf{user-item interaction} information, while other types of information without this symbol belong to items' \textbf{intrinsic properties}. The symbol $\heartsuit$ indicates the information is presented by \textbf{numerical} format, and the symbol $\spadesuit$ is for \textbf{descriptive} format, while the blank means the relevant information is absent.}
\label{datasets}
\end{table*}

A list of available datasets for \baby, along with their associated information types, is outlined in Table~\ref{datasets}. These datasets are selected from literature in SBR since its inception in 2016~\cite{GRU4Rec}. Additionally, we scrutinize works of general recommender systems in the last five years (2020--2025) and select datasets that are appropriate to our research topic. 

We can categorize various kinds of side information into the following two types: (1) interaction information, which refers to data generated during user-item interactions, such as interaction time, user rating or reviewing for an item, as marked with $^*$ in Table~\ref{datasets}; and (2) item intrinsic properties, which describe the inherent features of the item, including category, brand, price and so on.
Moreover, relying on different vehicles for conveying item features, various information can also be grouped into two categories: \textit{numerical information} of $\heartsuit$ and \textit{descriptive information} of $\spadesuit$, as illustrated in Table~\ref{datasets}. On one hand, the numerical information delivers the abstract meaning of an item using numbers, such as integer IDs for the category or real numbers for the item price. On the other hand, descriptive information characterizes an item via modality information like text or images, which can encompass rich semantics such as color, style and so on.
The details of these datasets, including contents, session formulation paradigms and related methods, are outlined as follows:

\begin{itemize}
    
    \item \textbf{Amazon}\footnote{\url{http://jmcauley.ucsd.edu/data/amazon/}} is scratched from the E-commerce website Amazon, capturing product reviews along with its rich metadata including category, price, and title/description text. This dataset covers various domains such as Beauty, Cellphones and Accessories, Toys and Games and more. Typically, data belonging to each domain is treated as a distinct dataset for recommendation tasks. Current models usually view user behaviors occurring within one day as a session in the Amazon dataset due to the time granularity in this dataset being precise to the day. Containing various kinds of information, the Amazon dataset is widely used in the community to evaluated recommendation algorithms~\cite{Bian@CIKM2023, Wu@ICDE2023, CoHHN, BiPNet, MMSBR}.
    % ~\cite{Bian@CIKM2023, Wu@ICDE2023}
    
    \item \textbf{Beer}\footnote{\url{https://cseweb.ucsd.edu/$\sim$jmcauley/datasets.html\#multi\_aspect}} contains user reviews about beer from Ratebeer and Beeradvocate websites. Besides the information listed in Table~\ref{datasets}, it also records sensory aspects of beer, such as taste, look, feel, and smell. The user-item interactions are presented in the granularity of second in this dataset, enabling the creation of sessions with fine-grained time spans, such as 20-second intervals. As far as we know, there is a few methods using this dataset to examine recommendation performance, \eg CoCoRec~\cite{cai@SIGIR2021}. 
    % ~\cite{cai@SIGIR2021} 

    \item \textbf{Cosmetics}\footnote{\url{https://www.kaggle.com/mkechinov/ecommerce-events-history-in-cosmetics-shop}} records behavior data over 5-month period spanning from October 2019 to February 2020 in a medium-sized cosmetics online store. Note that, this dataset explicitly highlights session IDs, eliminating the need for extra data partitioning for session formulation. Generally, data from each month is used as a separate dataset in existing studies~\cite{CoHHN,Chen@SIGIR2023,Chen@CIKM2023Short}.

    \item \textbf{Diginetica}\footnote{\url{https://competitions.codalab.org/competitions/11161}}, used in the CIKM Cup 2016 competition, consists of sessions extracted from e-commerce search engine logs. In this dataset, item title is presented via discreate token IDs instead of raw text. This dataset is popular in the task of SBR ~\cite{DPAN4Rec,song@CIKM2021,lai@SIGIR2022,Liu@NeurIPS2023}.

    \item \textbf{Foursquare}\footnote{\url{https://sites.google.com/site/yangdingqi/home/foursquare-dataset}} records digital footprints of users in location based social networks, crawled from Foursquare. It includes global-scale check-in data from restaurant venues across over 400 cities, covering a period of around 18 months, from April 2012 to September 2013. This dataset contains detailed geographic information about the restaurants, making it a valuable resource for studying point-of-interest (POI) recommendation, where the goal is to suggest places based on a user’s interests and location. In practice, data from different cities in this dataset are often used as separate benchmarks for evaluating recommendation models~\cite{Caser,Lian@KDD2020,Wang@ICDE2022,Qin@WSDM2023}.

    \item \textbf{Goodreads}\footnote{\url{https://mengtingwan.github.io/data/goodreads}} contains users' reviews towards books from the Goodreads website~\cite{Wan@RecSys2018}. This dataset records user-item interactions with minute-level granularity, allowing for the generation of sessions with arbitrary time intervals, such as 30 minutes. Additionally, since the dataset is originally presented in the form of a book graph, corresponding methods often utilize GNN to handle the recommendation tasks~\cite{Zhang@IJCAI2022,Ma@AAAI2020,Li@CIKM2021}.

    \item \textbf{Instacart}\footnote{\url{https://www.kaggle.com/c/instacart-market-basket-analysis}} describes grocery orders of users on the online grocery marketplace Instacart. Specifically, this anonymized dataset includes over 3 million grocery orders from more than 200,000 users. Owing to the vast volume of data within this dataset, it is a common practice to sample a subset from it for analysis and experimentation in existing methods~\cite{qin@SIGIR2021,DIMO}.

    \item \textbf{JingDong}\footnote{\url{https://tinyurl.com/ybo8z4yz}} captures various micro-behaviors of users, such as ordering, adding to cart, and commenting, from the Chinese e-commerce website JD.com. Besides recording the timestamp of user behaviors, this dataset also highlight the dwell time at the second level that users spend on each item. The dwell time can serve as an indicator of user preference, where the longer a user stays on an item, the stronger their interest in it~\cite{Alexander@arXiv2017}. Moreover, this dataset includes two item categories, \ie `Appliances' and `Computers', which are commonly used as separate datasets in related research~\cite{Huang@WSDM2019,meng@SIGIR2020,Yuan@ICDE2022}.

    % \item \textbf{KKBOX}\footnote{\url{https://www.kaggle.com/c/kkbox-music-recommendation-challenge/data}} is provided by a famous music service KKBOX, which contains users’ historical records of listening to music in a given period.

    \item \textbf{LastFM}\footnote{\url{http://ocelma.net/MusicRecommendationDataset/index.html}}, as a music-artist dataset, recording user interactions (listening) with music artists, collected from online music service Last.fm. In this dataset, user-provided tags for the artists are used as the category information. Due to its unique focus on music-related events, this dataset is particularly suitable for music interest recommendation~\cite{Zhou@CIKM2020,Huang@SIGIR2018,WangP@SIGIR2020,Ye@SIGIR2020}.

    \item \textbf{Movielens}\footnote{\url{https://grouplens.org/datasets/movielens/}} a classical rating dataset in the community, compiles user ratings about movies from the MovieLens website. The dataset is available in various sizes, each forming a distinct dataset used in different studies. For instance, MovieLens-100k and MovieLens-1m are commonly used variants, each named after the number of user-movie ratings they contain~\cite{liu@AAAI2021,Liu@TKDE2024,Xue@TOIS2023}.

    \item \textbf{NineRec}\footnote{\url{https://github.com/westlake-repl/NineRec}} consists of the user-watching behaviors on short videos from prominent short-video platforms in China, such as Bilibili, Kuaishou, and Douyin~\cite{NineRec}. This dataset is unique because it includes data from different channels within a single platform (e.g., vertical channels on Bilibili) as well as across different platforms (e.g., Bilibili versus Douyin). This feature makes the dataset particularly valuable for exploring both cross-domain and cross-platform recommendation scenarios~\cite{Fu@SIGIR2024,Li@ICDE2024}. As a newly released dataset, it has yet to be introduced and utilized in the field of SBR.

    \item \textbf{RetailRocket}\footnote{\url{https://www.kaggle.com/retailrocket/ecommerce-dataset}}, which is generated from the Retailrocket online shopping site, spans a period over 4.5 months and captures various kinds of user-item interactions, such as `view', `add to cart' and `transaction'. Note that, all words in text values were normalized and hashed in this dataset. Additionally, this dataset highlights the hierarchical relationships among categories using a category tree structure, which can be a valuable resource in recommendation research. It has been a popular benchmark used in various studies, including~\cite{Garg@SIGIR2019,NISER,yang@KDD2022}.

    \item \textbf{Steam}\footnote{\url{https://github.com/kang205/SASRec/tree/master/data}}, crawled from a famous online video game distribution platform Steam, contains user reviews about computer games from October 2010 to January 2018. This dataset includes unique information about video games, such as users' play hours, media scores and developer details. Such information can provide game manufacturers with valuable insights into user behaviors within the gaming domain. Some methods evaluate their recommendation performance under this game dataset~\cite{Li@WSDM2020,Zheng@AAAI2021,Seol@SIGIR2022Short,Zhang@KDD2023}.

    \item \textbf{Taobao}\footnote{\url{https://tianchi.aliyun.com/dataset/dataDetail?dataId=649}} captures users' online shopping behaviors from Taobao which is one of China's largest e-commerce platforms. This dataset is widely used in current research to address implicit feedback recommendation scenarios, providing valuable insights into user interactions and preferences~\cite{yang@KDD2022,Zhu@NeurIPS2019,Han@WWW2024,WW@TKDE2023,yuan@SIGIR2022}.

    \item \textbf{Tmall}\footnote{\url{https://tianchi.aliyun.com/dataset/dataDetail?dataId=42}} collects anonymized users' shopping logs over six months, including data before and on the `Double 11' day which is a major promotional event in Chinese e-commerce platforms. Originating from Tmall, one of the largest B2C platforms in China, this dataset includes the label information indicating whether a user is a repeat buyer for the same item. This feature is particularly valuable for studying repeat consumption behaviors. In addtion, this dataset was initially published for the IJCAI 2015 competition and has since been extensively used in related research~\cite{Hou@SIGIR2022Short,Han@SIGIR2022,zhou@WWW2022,Qiao@CIKM2023}.

    \item \textbf{Trivago}\footnote{\url{http://www.recsyschallenge.com/2019/}} is a session-based hotel recommendation dataset that records user actions during online hotel searches. Released by Trivago which is a global hotel search platform, this dataset was used in the RecSys Challenge 2019. It specifies the type of actions performed by users like view item rating, item information, item image and so on, providing valuable insights into user behaviors when booking hotels. Despite the richness of its user interaction data related to hotel bookings, it has been underutilized in academic research, with relatively few studies leveraging its potential~\cite{Yuan@ICDE2022}.

    \item \textbf{XING}\footnote{\url{http://2016.recsyschallenge.com}} collects user interactions with job postings, tailed specifically for job recommendation tasks~\cite{Du@AAAI2024,Francisco@RecSys2019}. Created by XING, a social network focused on job searching and recruitment, this dataset was used in the ACM RecSys Challenge 2016. It includes rich information about job postings, such as career levels (e.g., beginner, experienced, manager), industry types (e.g., Internet, Automotive, Finance), and employment types (e.g., full-time, part-time). Given the inherent sequential dependencies among user-job interactions, many efforts attempt to handle the job recommendation task with techniques in SBR~\cite{Wu@WWW2020,Kermany@WSDM2022Short,pang@WSDM2022,Zhang@TKDE2022}.

    \item \textbf{Yelp}\footnote{\url{https://www.yelp.com/dataset}} captures users' reviews for businesses across eight metropolitan areas from the online review platform Yelp.com. This dataset includes a wide range of business attributes, such as opening hours, parking availability, and ambience. Notably, its location information makes it well-suited for point-of-interest (POI) recommendation tasks. Moreover, due to its large size, many existing studies only utilize a subset of this dataset for their analyses and implementation~\cite{Tian@CIKM2022,Liu@CIKM2023Text,Chang@WSDM2024,FineRec}. 

    \item \textbf{Yoochoose}\footnote{\url{http://2015.recsyschallenge.com/challege.html}} contains six months of click-stream data from a European e-commerce website that sells a variety of items like garden tools, electronics and more. This dataset explicitly segments a user's behavior sequence into multiple sessions, making it a popular benchmark for SBR. Released in the RecSys Challenge 2015, Yoochoose is one of the earliest datasets designed for SBR and remains widely used in the domain~\cite{RepeatNet,choi@WSDM2022,Chen@SIGIR2022}.
    
\end{itemize}

\section{Data Characteristics and Utility}\label{dataUti}
The goal of side information-driven session-based recommendation is to incorporate various types of data to enhance user intent capturing. Specifically, each type of information possesses its unique characteristics and can shed light on different behavior patterns of users, thereby improving SBR from various aspects. For instance, `Time' information is instrumental in modeling sequential dependencies among items, whereas the `Price' factor can provide insights into user price sensitivity. Furthermore, these various types of data allow for the creation of distinct recommendation scenarios. Location data, such as the address information, is crucial for the formulation of point-of-interest recommendations. In the following contents of this section, thus, we will delve into the details of currently available side information. Specifically, we will explain the characteristics of each type information, highlight user behavior patterns that they help reveal, point out the specific recommendation scenarios that certain types of information could support, and finally discuss the collective relations of various information in affecting \baby. 

% In addition, we plan to summarize how these various data types are currently utilized in SBR solutions, discussing both the benefits and the challenges of these utilization paradigms. how each contributes to improving SBR

\subsection{Time Information}
Time information records the exact timestamps of user-item interactions, often represented in UNIX time format. This data can be easily converted into a detailed ‘year-month-hour-minute-second’ structure, offering a fine-grained view for user activities. In the context of SBR, time information plays a pivotal role in identifying the sequential dependencies among items, which is essential for capturing the evolving nature of user preferences~\cite{YangH@SIGIR2023,Li@SIGIR2022Short,Rashed@RecSys2022,Long@TKDD2024}. Moreover, the varying granularity of time data facilitates the detection of diverse sequential dependencies, which is of importance in interpreting user behaviors. For instance, by analyzing time data at the day-level granularity, it becomes intuitive to recommend quick meal options during weekdays when users may be pressed for time, and more leisurely dining options on weekends when they might prefer a more relaxed experience. Because it enables modeling of dynamic user intent, time information is of great importance in handling both SBR and sequential recommendation tasks.

\subsection{Category Information}
{Category} information refers to the hierarchical taxonomy used to classify and group items based on their types, characteristics, or other relevant attributes. In current datasets, category information could be represented either as integers or raw text, as illustrated in Table~\ref{datasets}. Typically, the item categories are organized as a tree structure, such as `Electronics' branching into subcategories like `Computer' and `Smartphones', with further subdivisions like `Laptops' and `Desktops' under the `Computer' branch. This hierarchical structure mirrors the users' decision-making process, as they progressively narrow down their intentions, such as navigating from `Clothes' to `Men's Clothing' and finally to `Pants'.  Leveraging these category hierarchies can enhance recommender systems by significantly reducing the number of candidate items, leading to more accurate recommendations. For instance, suggesting phone cases from the `phone-accessories' subcategory would be highly relevant for users who recently bought a smartphone from the `Smartphones' category.

% Unfortunately, effectively exploring the informative hierarchical relationships behind item categories remains a challenge. Existing methods typically resort to one-hot encoding to transform item categories, with numerical or descriptive format, into inputs suitable for neural models~\cite{cai@SIGIR2021,xie@SIGIR2022}. While this approach simplifies category data for model processing, it fails to retain the hierarchical relationships inherent in the category structure. This results in a significant loss of valuable contextual information.

\subsection{Brand Information}
Brand information encompasses details about the brand or manufacturer of an item, reflecting intangible item attributes, such as reputation or quality. It serves as a unique indicator of manufacturer state among the currently available information. Similar to item categories, brand information is typically represented as integers or raw text. The brand of an item plays an unique role in guiding customers' purchasing decisions. Specifically, the `brand effect', widely studied in economics, underscores how brand loyalty influences a consumer's behaviors. Recommender systems can harness this effect to deliver more personalized and relevant suggestions. For example, users who exhibit a strong preference for a particular brand are likely to respond favorably to recommendations featuring products from that brand. This alignment with user preferences can significantly enhance the effectiveness of recommendation strategies, making `Brand' information a valuable asset to meet individual tastes. 

% Despite its potential impact, none of the existing methods has specifically focused on this interesting phenomenon. This oversight leads to a missed opportunity in harnessing the useful brand information.

\subsection{Price Information}
{Price} denotes to the monetary cost assigned to an item, representing the amount a user is required to pay for its acquisition. In current datasets, item prices are commonly presented with the format of real numbers, such as \$19.99.
An item's price is undeniably a key factor influencing a user’s decision when selecting products~\cite{Wu@ICDE2022,Zheng@ICDE2020,Wan@WWW2017price}. Typically, users approach shopping with a predetermined budget or price range in mind. No matter how interested they may be in a particular item, if its price exceeds their budget, they are likely to abandon the purchase, emphasizing the importance of aligning recommendations with users' financial expectations. Owing to its critical influence on user decisions, the integration of item price into SBR task holds significant potential to boost recommendation effectiveness and its applicability in real-world scenarios. By taking price factors into considerations, SBR models can offer recommendations that are not only aligned with user interest but also their financial expectations, thereby greatly improving user satisfaction~\cite{CoHHN,BiPNet}. 

% Although promising, effective implementation of item price into SBR is a non-trivial task duo to following challenges: (1) the price range of items can vary significantly, even by several orders of magnitude, making it difficult to apply a unified standard to encode the item price for its utilization in practice; and (2) users' price preferences are dynamic instead of static, varying greatly along with many factors like the seasonal factors (users might be more inclined to spend more during holiday seasons). 

\subsection{Text Information}
{Text} information provides detailed descriptions of items using natural language, typically presented in the form of raw words. As outlined in Table~\ref{datasets}, we primarily focus on two common types of text: item \emph{title} and \emph{description}. The item title serves as a concise descriptive snippet, summarizing the item with a focus on its key features or selling points, such as size, capacity, and material. In contrast, the description offers a more comprehensive overview, giving potential buyers detailed insights into aspects like usage instructions and support information. With rich semantics, these textual elements serve as a valuable resource for enhancing recommender systems. On the one hand, they help articulate item features, such as highlighting Marvel-related factors in \fig~\ref{intro}. On the other hand, they can reveal users' fine-grained preferences, like identifying Marvel fans in \fig~\ref{intro}. Moreover, with achieving tremendous success across various tasks, Large Language Models (LLMs) have been widely applied in the field of recommender systems~\cite{Wang@AAAI2024,Lin@SIGIR2024,Zhao@SIGIR2024}. Obviously, text information is the foundation in the development of LLM-based recommender systems. Furthermore, text also plays a vital role in advancing research within multi-modal recommender systems~\cite{Hu@CIKM2023,Zhou@WWW2023,wei@MM2020}.

% A common practice involves extracting text semantics using pre-trained models, like BERT~\cite{hou@KDD2022,Liu@CIKM2023Text}, with the aim of enhancing model performance by incorporating these semantics.
% However, a notable challenge arises from the fact that these pre-trained models are typically trained on general datasets, which may not align well with the context of recommendation. 
% As a result, they might struggle to extract relevant and useful semantics essential for recommendation task.  In addition, the recommendation effectiveness can vary considerably based on the scale of pre-trained models~\cite{Yuan@SIGIR2023}.This variability makes it difficult to determine whether a model's success is due to its own design or the advanced capabilities of the pre-trained models employed, impeding fair comparison.

\subsection{Image Information}
{Image} information refers to the visual representations of items, commonly in the form of digital photographs or graphics. These images provide users with a clear and intuitive understanding of what the items look like, effectively showcasing key features such as color and style. Visual content plays a unique role in reducing the uncertainty that often accompanies online shopping, where users cannot physically inspect the products. By harnessing the detailed insights gained from images, recommender systems are able to offer more precisely tailored suggestions, matching closely to individual user interests. As depicted in \fig~\ref{intro}, item images can be leveraged to identify a user's specific interests on Marvel-themed items, thereby significantly enhancing the effectiveness of recommendations. Similar to text information, item images are a vital resource supporting the research on multi-modal recommendation scenarios~\cite{MMSBR,Geng@EMNLP2023,Liu@TOIS2023}.

\subsection{Address Information}

Address information signifies the geographic location of certain items such as restaurants, tourist attractions, or other points of interest. This information is typically provided in varying levels of granularity, ranging from precise coordinates (latitude and longitude) to broader regions like`Northeast' or `Southwest'. Address information is crucial in influencing user decisions, especially on location-based services including location-based advertising and online food delivery~\cite{Qin@WSDM2023}. For instance, under restaurant recommendations, the proximity of a restaurant plays a significant role in a user’s decision to visit since that users generally prefer dining at places that are conveniently located nearby. Moreover, address information forms the basis for the specialized task of point-of-interest (POI) recommendation, where the goal is to suggest new places, such as restaurants or tourist spots, tailored to a user’s preferences and location~\cite{Ma@CIKM2018,Wang@ICDE2022,Qin@WSDM2023}.

\subsection{Rating Information}
{Rating} information serves as a quantifiable indicator of a user's overall attitude toward an item, typically expressed on a scale from 0 to 5. Generally, higher ratings indicate a stronger preference or satisfaction with the item. Unlike other types of information provided by manufacturers, ratings are uniquely generated by users themselves. This user-generated nature makes ratings particularly valuable for crafting personalized recommendations. For instance, by analyzing users' explicit ratings across various items, recommender systems can identify and suggest items similar to those that have received high ratings, while avoiding items with lower ratings. The incorporation of ratings aligns recommendations more closely with the user's demonstrated preferences, resulting in a more tailored and satisfying user experience. Additionally, rating information is central to the classical task of rating prediction, which was a primary focus in the early development of recommender systems~\cite{Tang@RecSys2013,Wang@IJCAI2016}.

% However, this user-generated nature of ratings introduces unique challenges for its implementation. Users with diverse preferences may rate the same item very differently, leading to potentially conflicting ratings.  This makes it challenging for recommender systems to accurately evaluate and represent an item under such complex situations. 

\subsection{Review Information}
Review information consists of written feedback provided by users who have purchased and experienced the items. Unlike ratings, which offer a numerical summary of user satisfaction, reviews delve into the specifics of user preferences, item features, and overall experiences through raw text. This detailed content offers valuable insights that can significantly enhance the precision and relevance of recommendations. For instance, if a user mentions a preference for loose-fitting clothing in a review, the system can prioritize recommending items that match this preference, avoiding the suggestion of unsuitable tight-fitting options. Besides, reviews allow for an evaluative perspective on the item itself. Specifically, if the majority of reviews criticize an item for poor quality, it suggests that the item might not be suitable for users seeking high-quality products. By integrating these insights from reviews, recommender systems can deliver more reasonable and user-aligned recommendations. In addition, the rich, descriptive content found in reviews serves as a valuable resource for LLM-based approaches to analyze and understand user preferences at a deeper level.

% Nevertheless, effectively leveraging reviews in SBR poses significant challenges. On one hand, user-item reviews tend to be colloquial, leading to the difficulty to extract relevant semantics form them. On the other hand, an item can receive a wide range of reviews from different users, with opinions that may vary greatly or even be contradictory.  Similar as ratings,  these diverse perspectives add complexity to the task of accurately representing an item.

\subsection{Behavior Information}
{Behavior} information captures the specific types of user-item interactions, such as clicks, add-to-cart, or purchases, typically presented in raw text format. Diverse types of behavior offer distinct insights into users' intents and preferences regarding items. For instance, a user clicking on an item may not necessarily indicate a strong liking. Instead, it could just be a part of casual browsing. In contrast, a behavior such as making a purchase is a strong indicator of user satisfaction with the item. The consideration of these diverse behaviors allows for a more fine-grained and deeper interpretation of user-specific preferences, thus holding the potential for more effective recommendations. This focus on different user behaviors has also given rise to a new research topic known as `multi-behavior recommendation' whose aim is to enhance user intent modeling by integrating various types of user interactions~\cite{Li@WSDM2024,Yan@SIGIR2024, Guo@TKDE2025}.

% Reflecting user specific intentions, behavior types have receive increasing attention in recent SBR research,  giving rise to a new sub-topic known as ``Multi-behavior''  SBR. The main challenges in the Multi-behavior SBR lie in (1) discerning the varying importance of different behaviors in reflecting user interest; (2) capturing user personalized behavior patterns, where some users might directly purchase items they favor, while others may prefer to add items to their shopping cart for collective purchase later.  

\subsection{Discussion}
Beyond their individual effects, various types of side information also exhibit collective relations in user behavior modeling within \baby, which can be either complementary or conflicting. On one hand, different types of information can be complementary in capturing user behavior patterns. For example, users may be more willing to accept higher prices for well-known brands, making the joint consideration of item price and brand valuable for understanding their price preferences~\cite{CoHHN, BiPNet}. Similarly, ratings quantify user satisfaction, whereas reviews provide explanations and sentiment. Their fusion may improve understanding of user preferences beyond numerical scores. On the other hand, conflicts may arise among different types of side information. For instance, image and text descriptions of an item do not always align, \eg misleading promotional images or exaggerated textual descriptions can introduce inconsistencies, making it challenging to extract coherent item representations~\cite{MMSBR}. Overall, effectively modeling these relations requires careful design, like modality alignment techniques or adaptive fusion mechanisms, to resolve conflicts and leverage complementary signals.

\section{Usage of side information}\label{Usage}

Side information-driven session-based recommendation is centered around the strategic use of diverse data to deliver personalized suggestions catering to user preferences. In previous sections, we elaborated on the characteristics of various side information and their utility in revealing user intents—essentially addressing what side information is and why it is valuable. This section shifts the focus to the practical aspect of side information, \ie how to use these side information. Drawing from a comprehensive review of \baby, we plan to detail the usage of side information across three key areas: (1) Data encoding, \ie the pre-processing stage, where we explain how to encode various side information to serve as inputs for subsequent neural models; (2) Data injection, an overview of manners for integrating side information into the recommendation task; and (3) Involved techniques, a summary of common techniques used in \baby, such as RNN, attention mechanism, GNN and so on.

\subsection{Data Encoding}
As discussed in Section~\ref{benchmark}, we classify various kinds of side information into two groups, \ie numerical information and descriptive information, based on their vehicles for portraying item features. In the meantime, these two types of information are typically encoded in different ways by current neural models in \baby. 

For numerical information, the data is first converted into consecutive integers and then mapped into dense embeddings using a look-up embedding table~\cite{CoHHN,song@CIKM2021,Liu@RecSys2023}. This transformation allows raw numerical values to be seamlessly integrated into neural networks. However, recent methods have recognized that original numerical information may contain unique relationships, such as the hierarchical structure of item categories. To preserve these relationships during the integer transformation, some approaches have been developed. For instance, CoHHN~\cite{CoHHN} and BiPNet~\cite{BiPNet} have observed that item price distribution often follows a logistic distribution, so they convert item price into discrete price levels based on this distribution instead of commonly used uniform one. This ensures that the encoded numerical data preserves its intrinsic characteristics, enabling the generation of more meaningful and informative representations.

In contrast, descriptive information, such as text and images, requires more sophisticated processing techniques. Specifically, existing approaches tend to employ advanced pre-trained models from Natural Language Processing (NLP) and Computer Vision (CV) to represent this type of information~\cite{Wei@MM2019,Sun@CIKM2020,Wang@TOIS2024}. For example, text data might be processed using models like Glove~\cite{Glove} or BERT~\cite{BERT}, which are capable of generating rich text embeddings based on their extensive training on massive language corpora. Similarly, image data can be represented using models like GoogLeNet~\cite{googlenet} or ViT~\cite{ViT}, which are well-trained on vast image datasets and are highly effective at capturing intricate visual features. These pre-trained models are equipped with rich knowledge about their respective modalities, allowing them to generate robust embeddings for text and image data. By leveraging these embeddings, recommender systems can enhance the quality and relevance of their recommendations.

\subsection{Data Injection}
The key points of neural SBR lie in two aspects: item representation learning and session representation learning. Once these embeddings are established, recommendation lists can be generated based on the similarity between them. Relying on a single type of item ID information, conventional methods in SBR usually resort to a straightforward manner to obtain item ID embeddings and aggregate them into session embeddings to achieve personalized suggestions. However, the complexity increases significantly in side information-driven SBR due to the involvement of the utilization for various kinds of side information. Obviously, how to effectively inject these types of side information into neural models possesses huge impact on recommendation performance. Unfortunately, the injection of side information in SBR has received little attention in current research. 

\begin{figure*}[t]
  \centering
  \includegraphics[width=0.98\linewidth]{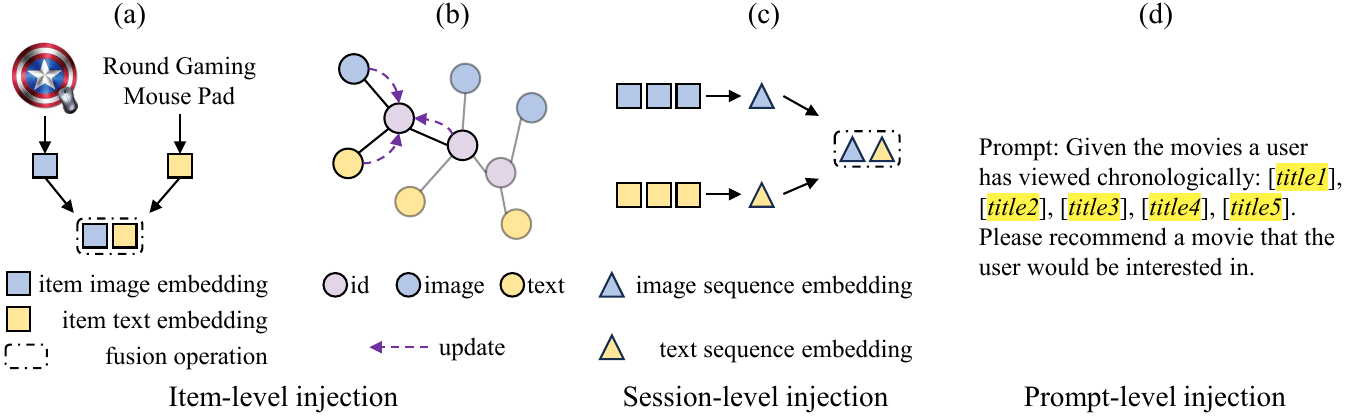}
  \caption{Three manners for side information injection: item-level injection, session-level injection and prompt-level injection}\label{injection}
% \vspace{-0.1in}
\end{figure*}

To this end, we summarize the main manners of injecting side information within \baby into three types, including item-level, session-level and prompt-level injection, as illustrated in~\fig~\ref{injection}. The first two methods are commonly applied in neural models and are organized based on when the injection occurs within the aforementioned workflow of \baby, \ie Item encoder, Behavior modeling, Session encoder and Recommender. The last method, prompt-level injection, is particularly relevant for approaches using Large Language Models (LLMs). In addition, we will discuss the limitations associated with each type of side information injection.

\subsubsection{Item-level Injection}

Item-level injection incorporates various kinds of side information into SBR on item representation learning, specifically occurring during the Item encoder and Behavior modeling stages, \ie the first two steps of standard workflow in \baby.

In practice, item-level injection first combines various item side information into unified item embeddings. These unified embeddings then serve as the foundation for representing an anonymous user's intent within a session~\cite{lai@SIGIR2022,Hu@CIKM2023,liu@AAAI2021,Liu@RecSys2023}. For instance, item encoder converts an item's image and text into dense embeddings, respectively. These embeddings can then be merged into a unified item embedding using simple concatenation~\cite{FGNN,Shalaby@RecSys2022,lai@SIGIR2022,liu@WWW2020}, as illustrated in~\fig~\ref{injection} (a). Another common approach for item-level injection involves the use of Graph Neural Network (GNN). As presented in~\fig~\ref{injection} (b), some works~\cite{CoHHN,Yuan@ICDE2022,BiPNet,Wu@ICDE2023} create a heterogeneous graph with nodes representing different types of side information. Through performing node embedding updating on this graph, unified item embeddings integrating various side information can be learned. 

Although intuitive and straightforward, it is non-trivial to conduct effective item-level side information injection. At first, the inherent heterogeneity and noise across different types of information make it challenging to obtain robust embeddings for these information~\cite{MMSBR,Tadas@TPAMI2019, bottleneck}. In addition, this item-level injection incorporates various types of information before modeling user behavior sequences. Consequently, it falls short of capturing user preferences specific to each type of information. Taking the important price factor as an example, when price is blended with other information at the item level, the model may struggle to discern a user's price sensitivity. This lack of clarity impedes the model's ability to accurately understand user behaviors, which in turn negatively impacts the overall recommendation performance.

% \noindent\textbf{Item first manner}. On one hand, it is straightforward that firstly integrating various side information into a unified item embedding, and then aggregating item embeddings as conventional SBR does to represent a session~\cite{lai@SIGIR2022,Hu@CIKM2023,liu@AAAI2021,Liu@RecSys2023}. Although intuitive, it is non-trivial to conduct effective item side information fusion due to the inherent heterogeneity and noise (as stated before) existing in the multiple types of information. Moreover, this paradigm merges various types of information before modeling user behavior sequences. Consequently, it falls short of capturing user preferences specific to each type of information. Taking the important price factor as an example, blending it with other information at item-level makes it difficult for the model to perceive user price sensitivity, impeding the accurate understanding for user behaviors.

\subsubsection{Session-level Injection}
Session-level injection relies on session representation learning to introduce side information into \baby, where the side information injection is conducted in Session encoder and Recommender stages.

In this injection manner, neural models prioritize user behavior sequence over a single action (\ie~item). Specifically, they first build sequences for each type of information within the session, then represent each sequence independently, and finally generate recommendation list by considering all sequence embeddings~\cite{Zhou@CIKM2020,song@CIKM2021,MMMLP}. As illustrated in ~\fig~\ref{injection} (c), for a session, its item image sequence and item text sequence, which have been processed by Behavior modeling separately, are converted into image and text sequence embeddings. Afterwards, these sequence embeddings are combined, like using concatenation, to represent user overall preferences for recommendation generation~\cite{zhang@IJCAI2019,WW@TKDE2023}.

This paradigm holds the merit of capturing users' specific preferences related to certain types of information. However, lacking in fusing side information at the item level, they struggle to obtain informative item embeddings. Another challenge lies in synthesizing a comprehensive understanding of users' overall preferences based on their unique interests in each type of information. In reality, users' final decisions are typically influenced by a combination of various factors. Unfortunately, there are not explicit signals to indicate how these factors interrelate to drive user actions, making it difficult to perform effective predictions in this manner. 

% \noindent\textbf{Session first manner}. On the other hand, some efforts prioritize user behavior sequence over a single action (\ie~item). They build multiple sequences based on various types of information, represent each sequence independently, and then merge the embeddings of these sequences to capture a user's overall interest~\cite{CoHHN,song@CIKM2021,MMMLP}. This paradigm holds the merit to capture users' specific preferences related to certain factors. However, lacking for fusing side information at item level, they struggle to obtain informative item embeddings. In addition, how to obtain users' comprehensive preferences based on these unique interest on each type information is also a thorny problem. Generally, users' final decisions are influenced by a combination of these diverse interests. Unfortunately, there are not explicit signals to indicate how these factors interrelate to drive user actions, making it challenging to perform effective predictions in this manner. 

\subsubsection{Prompt-level Injection}\label{prompt_injection}
Prompt-level injection is uniquely designed for LLM-based methods in \baby (refer to Section~\ref{llm} for more details on LLM-based \baby), distinguishing itself from item- and session-level injection tailored  for neural models. A prompt acts as a structured template containing raw text, designed to trigger an LLM's ability to complete a specific task~\cite{liu@CS2023,Lin@SIGIR2024,Zhao@SIGIR2024}. Generally, a prompt consists of a task description, an instruction, and the side information to be injected. For instance, as illustrated in~\fig~\ref{injection} (d), prompt-level injection may involve placing item titles into the template (i.e., injecting side information into the task context) and then leveraging the extensive knowledge encoded within the LLM to generate personalized recommendations.

Despite facilitating the application of LLM on recommendation tasks, the prompt-level injection still faces several challenges that limit the full potential of LLMs in this domain. Firstly, a prompt based solely on a single user’s behavior lacks the collaborative knowledge derived from the collective actions of a large user base. This collaborative knowledge is crucial in recommendation systems for identifying shared preferences and trends among users~\cite{DIMO,Yuan@SIGIR2023}. Furthermore, designing an effective and universally applicable prompt template is difficult, as subtle differences in template structure can lead to significant variations in LLM's performance. Lastly, prompt-level injection is uniquely suitable for text information, making it challenging to integrate other types of data, such as images or numerical attributes. This limitation restricts the versatility and effectiveness of LLM in recommendation domain.

\subsection{Involved Techniques}
Various neural structures have been leveraged in \baby, including RNN, attention mechanisms, GNN, LLM, and others. These techniques are utilized in the behavior modeling stage to handle user actions within sessions. This section will examine these techniques from several perspectives: formulations, unique contributions to uncovering user preferences, representative models that instantiate these techniques, and the limitations associated with each technique. 
% By doing so, we aim to provide a comprehensive overview of how these techniques contribute to advancing \baby and where they may face challenges. 
Note that, we exclude the machine learning based methods in SBR, such as matrix factorization~\cite{BPR-MF}, markov chain~\cite{MDP} and nearest-neighbor-based approaches~\cite{IKNN}, as these methods focus on modeling item ID sequences rather than incorporating side information for performing SBR.

\subsubsection{Recurrent Neural Networks (RNN)}
RNN is one of the earliest neural structures applied in SBR~\cite{GRU4Rec}. Given its inherent ability to model ordered data, RNN-based methods are able to capture sequential dependencies among user actions. Taking the representative Gated Recurrent Unit (GRU) as an example, we can formulate its processing as follows,

\begin{equation}
    \begin{split}
         \textbf{h}_j &= ( 1 - \textbf{z}_j) \textbf{h}_{j-1} + \textbf{z}_j \hat{\textbf{h}_j}, \\
         \textbf{z}_j &= \sigma ( W_z \textbf{v}_j + U_z \textbf{h}_{j-1} ), \\
        \hat{\textbf{h}_j}  &= tanh[ W \textbf{v}_j + U ( \textbf{r}_j \odot \textbf{h}_{j-1} )], \\
        \textbf{r}_j &= \sigma ( W_r \textbf{v}_j + U_r \textbf{h}_{j-1} ),
    \end{split}
\end{equation}
where $ W, W_r, W_z, U_r, U_z $ are weight matrices; $ \textbf{v}_j $ is the embedding of $j$-th item in the session; $ \textbf{h}_j \in R^k $ denotes the output of $j$-th hidden state in GRU layer ($ \textbf{h}_0 $ = 0); and $ \sigma $ denotes the sigmoid function.

% The initial state of the GRU is set to zero vectors, i.e., $ h_0 $ = 0.

In the early stage of SBR's development, RNN and its variants are popular choices for handling side information in the task. For example, MV-RNN~\cite{MV-RNN} and TMRN~\cite{Huang@WSDM2019} rely on item-level injection to incorporate side information. P-RNN~\cite{P-RNN} explores both item-level and session-level injection methods, attempting to capture the contextual impact of side information at different granularities. 
% For instance, models like P-RNN~\cite{P-RNN} and TMRN~\cite{Huang@WSDM2019} used Gated Recurrent Unit (GRU), while MV-RNN~\cite{MV-RNN} employed Long Short-Term Memory (LSTM). 
Nowadays, RNN mostly severs as the basic component for deriving session embeddings, and is combined with other techniques, like GNN~\cite{FGNN,Yuan@ICDE2022} and attention mechanism~\cite{liu@WWW2020}, to generate final suggestions.

The limitations of RNN in \baby are notable: (a) The assumption of strict order between any successive items in RNN models is not always valid in real-world scenarios ~\cite{Wang@IJCAI2019, MCPRN}; and (b) Inherent recursive natural of RNN makes it hard to parallelize, leading to high time consumption and poor efficiency~\cite{STAMP,DIDN,BERT4Rec}.

\subsubsection{Convolutional Neural Networks (CNN)} CNN is another pioneering neural structure used in the early stage of SBR~\cite{Tuan@RecSys2017, Caser}. Generally, We can formulate the processing of GNN as follows,
\begin{align}
    \mathbf{o} &= \text{Conv}([\mathbf{v}_1;\mathbf{v}_2; ...;\mathbf{v}_t]),
\end{align}
where [;] denotes concatenation operation, Conv(*) is the convolutional layer including stacked filter and pooling operations, and $\mathbf{o}$ is the output, which is expected to encapsulate the local patterns of user actions.

Compared with the widely used RNN, CNN offers three main advantages. Firstly, CNN does not impose a strict order assumption on successive items in sessions, enabling more robust recommendation performance. Secondly, CNN is inherently designed for parallel processing, improving model efficiency. Lastly, CNN is able to effectively learn local patterns within small groups of items in a session, facilitating understanding of item relationships. 

Despite these benefits, CNN has seen limited applications in SBR. Notable examples include Caser~\cite{Caser} and NextItNet~\cite{Yuan@WSDM2019}. Besides, only a few efforts, like 3D-CNN~\cite{Tuan@RecSys2017}, have attempted to handle side information via CNN. In this approach, an item's embeddings of various side information are stacked into a matrix and fed into convolutional layers for side information incorporation. 

One of the main challenges with CNN in this context is the need to determine an optimal filter size, which indicates the range of local patterns CNN can capture. Obviously, it is hard to determine optimal range in practice, thereby significantly limiting CNN's broader applications in the domain of \baby~\cite{LINet}.

\subsubsection{Attention Mechanisms}
Attention mechanism is good at discerning the importance of distinct tokens within a sequence, making it particularly useful for SBR. In this context, attention mechanism is typically employed to highlight interactions that are relevant to user preferences while downplaying less relevant ones, thereby enhancing the learning of user interests~\cite{SASRec,STAMP,BERT4Rec}. We can formulate the attention mechanism via representative SASRec~\cite{SASRec} as follows,
\begin{equation}
    \begin{split}
         \textbf{o} = softmax(\frac{QK^{\top}}{\sqrt{d}})V, \\
         Q = W^Q[\textbf{v}_1; \textbf{v}_2; ...; \textbf{v}_t],\\
         K = W^K[\textbf{v}_1; \textbf{v}_2; ...; \textbf{v}_t],\\
         V = W^V[\textbf{v}_1; \textbf{v}_2; ...; \textbf{v}_t],
    \end{split}
\end{equation}
where $W^Q$, $W^K$ and $W^V$ are used to project inputs into Query, Key and Value spaces, respectively.

Currently, attention mechanism is widely adopted for analyzing user behaviors on various side information. For instance, NOVA~\cite{liu@AAAI2021} integrates side information by using item IDs as the Value and side attributes as the Query and Key. M2TRec~\cite{Shalaby@RecSys2022} and MMSBR~\cite{MMSBR} first merge side information into unified item embeddings, and capture user intents via transformer architecture.  In contrast, CBML~\cite{song@CIKM2021} handles different types of side information separately, and performs information fusion at session-level.

However, attention-based models depend heavily on additional position information to capture sequential dependencies among items~\cite{ludewig@UMUAI2020,Wang@CS2022}. The effectiveness of these models is highly sensitive to the schema of position encoding, leading to their unstable performance~\cite{Qiu@TOIS2021}.

\subsubsection{Graph Neural Networks (GNN)}
GNN has become a dominant neural structure in SBR recently. Its strength lies in effectively capturing complex transition relationships among items within sessions, leading to the generation of informative item embeddings. Generally, GNN-based methods operate on a specially constructed graph $\mathcal{G}$ derived from session data, updating an item embedding ($\textbf{v}_j$) via,
\begin{equation}
    \textbf{v}_j^* = g(\textbf{v}_j, f_{x_k \in \mathcal{G}_{j}}(\textbf{v}_k)),
\end{equation}
where $\mathcal{G}_{j}$ is a node set containing nodes adjacent to $x_j$ within the $\mathcal{G}$, $f(*)$ is used to extract relevant semantic information from these neighboring nodes, and $g(*)$ updates the embedding of target node via aggregating the extracted semantics. 

Under \baby, existing methods usually resort to heterogeneous graphs to capture diverse types of side information such as MKM-SR~\cite{meng@SIGIR2020}, FAPAT~\cite{Liu@NeurIPS2023}, REKS~\cite{Wu@ICDE2023} and GHTID~\cite{Li@WSDM2024}, using item-level injection as shown in \fig~\ref{injection}(b). Besides, some efforts extend general graphs to hypergraphs with degree-free hyperedges for capturing high-order relations among different kinds of nodes, as seen in models like CoHHN~\cite{CoHHN}, BiPNet~\cite{BiPNet} and MBHT~\cite{yang@KDD2022}. 

Despite their significant contributions to advancing \baby, GNN-based methods suffer from some limitations, such as the need for manually constructing graphs and relying on heuristic aggregating algorithms.  These limitations pose major obstacles to their practical implementations, restricting their broader applicability in real-world cases. 

% \subsubsection{Hypergraph Neural Network (HNN)}

% \subsubsection{Multi-Layer Perceptron (MLP)}

\subsubsection{Contrastive Learning (CL)}
CL, as a classical paradigm of self-supervised learning, has been a potent remedy for alleviating the inherent data sparsity issue in SBR~\cite{Xia@CIKM2021,Xie@ICDE2022}. Typically, contrastive loss of CL-based methods can be formulated as follows,
\begin{align}
    % \begin{split}
        \mathcal{L}_{con} = -\sum \frac{<\mathbf{r}, \mathbf{r}^{+}>}{<\mathbf{r}, \mathbf{r}^{+}> + \sum_{k}<\mathbf{r},  \mathbf{r}_{k}^{-}>},
    % \end{split}
\end{align}
where $<*>$ is similarity function, $\mathbf{r}^{+}$ and $\mathbf{r}_{k}^{-}$ are positive and negative samples about $\mathbf{r}$, respectively. CL-based methods typically obtain positive sessions by cropping, masking, or reordering certain items within a session, while viewing other sessions in the same batch as negatives. Subsequently, they are able to refine session embeddings via pulling positives closer and pushing negatives apart. 

In the context of \baby, current CL-based methods often leverage sequences from different types of side information to construct positives, such as in S$^3$-Rec~\cite{Zhou@CIKM2020}, HPM~\cite{Huang@SIGIR2023}, MSSR~\cite{Lin@WSDM2024} and Ti-model~\cite{Dang@AAAI2023}. 

The primary limitation of this technique lies in its inability to guarantee the generation of robust positives and negatives. Concretely, it is challenging to generate positives with truly similar semantics or negatives with clearly dissimilar meanings, leading to less effective self-supervised learning.

\subsubsection{Large Language Models (LLM)}\label{llm}
LLM, known for its extensive world knowledge, is an emerging technique in the domain of Natural Language Processing (NLP), achieving impressing performance in various tasks. Generally, the current usage of LLM in recommendation tasks can be summarized into two manners: LLM as an assistant and LLM as a recommender. In the first approach, relevant methods utilize well-trained LLM to extract specific information from raw data to improve their understanding of user intent. For instance, FineRec~\cite{FineRec} employs a LLM to process user-item reviews and extract item attributes that users care about. In the other approach, LLMs are directly trained using fine-tuning, prompt tuning or instruction tuning  based on recommendation data~\cite{Hu@WWW2024,Zheng@WWW2024,Sun@SIGIR2024}. Specifically, these methods focus on creating informative prompts, as discussed in prompt-level injection of Section~\ref{prompt_injection}, to evoke LLM's capability in generating personalized suggestions. 

Despite opening up a promising avenue to handle SBR, LLM still suffers from several obstacles that hinder their effective applications in the task. Firstly, while LLMs are trained on vast datasets from general domains, they lack the specific knowledge required for recommendation tasks, \eg collaborative knowledge, leading to their limitations in accurately understanding user behaviors. Secondly, the hallucination issue poses a significant challenge, where LLM-generated recommendations can sometimes deviate from reality. Although this may occur infrequently, it can negatively impact users' experience and diminish their trust in the system. Lastly, with a massive number of parameters, often in the billions, LLM's implementation in recommendation scenarios poses challenges related to both efficiency and effectiveness, making practical deployment difficult.

\section{Research Progress}\label{ResearchP}
% With substantial potential in revealing user intents, side information has received increasing attention in SBR recently. 
% The types of information used in current methods are crucial, as they indicate specific aspects of user behaviors the methods plan to capture. For instance,  `Time' information is instrumental in modeling the sequential dependencies among items, whereas the `Price' factor can provide insights into user price sensitivity. Besides, the information type also dictates the data processing approach and influences the formulation of datasets. 
With substantial potential in revealing user intents from various perspectives, side information has received increasing attention in SBR recently. As discussed previously, the types of side information are of great importance in guiding the design and implementation of recommendation algorithms. To this end, we introduce a new taxonomy that places particular emphasis on the type of side information to examine research progress within \baby. Our objective with this novel taxonomy is to offer a unique and insightful reference for future research, enabling researchers and practitioners to conveniently and efficiently identify, replicate, and evaluate algorithms that focus on specific types of side information. This unique approach not only facilitates a comprehensive understanding of existing methods but also highlights areas for potential advancements related to various information in the field. In the subsequent sections, we will outline the representative methods categorized by this taxonomy and point out their limitations. For clarity, Table~\ref{approach} offers a comprehensive summary of representative models within \baby, detailing the publication venues, types of side information integrated, and datasets utilized in these approaches.

\begin{table*}[!th]
\small
\tabcolsep 0.03in 
\caption{An overview of the side information-driven session-based recommendation approaches.}
  \centering

\begin{tabular}{c|c|c|c|    c|c|c|c|    c|c|c|c}
\toprule
\multirow{2}*{Models} & \multirow{2}*{Venues} & \multicolumn{9}{c|}{Types of Incorporated Information} & \multirow{2}*{Datasets} \\\cline{3-11}
& & {Category}    & {Brand}   & {Price}   & {Text}  & {Image} & {Address}   & {Rating}  & {Review}  & {Behavior} & \\\hline

{P-RNN~\cite{P-RNN}}    & {RecSys16'}    &{}    & {}   & {}   & {$\checkmark$}  & {$\checkmark$}   & {}   & {}  & {}  & {}    & {Private} \\  
 \hline
 {3D-CNN~\cite{Tuan@RecSys2017}}    & {RecSys17'}    &{$\checkmark$}    & {}   & {}   & {$\checkmark$}  & {}   & {}  & {}  & {}  & {}    & {Private} \\  
 \hline
  {DER~\cite{Chen@AAAI2019}}    & {AAAI19'}    &{}    & {}   & {}   & {}  & {}  & {}  & {}    & {$\checkmark$}  & {}    & {Amazon,Yelp} \\
\hline
{FDSA~\cite{zhang@IJCAI2019}}    & {IJCAI19'}    &{$\checkmark$}    & {$\checkmark$}   & {}   & {}  & {}   & {}  & {}    & {}  & {}    & {Amazon, Tmall} \\  
 \hline
{RNS~\cite{RNS}}    & {IJCAI19'}    &{}    & {}   & {}   & {}  & {}  & {}  & {}    & {$\checkmark$}  & {}    & {Amazon} \\  
 \hline
 {TMRN~\cite{Huang@WSDM2019}}    & {WSDM19'}    &{$\checkmark$}    & {}   & {}   & {}  & {}   & {}  & {}  & {}  & {}    & {Amazon, JingDong, LastFm} \\  
 \hline
 
 {S$^3$-Rec~\cite{Zhou@CIKM2020}}    & {CIKM20'}    &{$\checkmark$}    & {$\checkmark$}   & {}   & {}  & {}   & {}  & {}  & {}  & {}    & {Amazon, LastFm, Yelp} \\  
 \hline
 {GeoSAN~\cite{Lian@KDD2020}}    & {KDD20'}    &{}    & {}   & {}   & {}  & {}   & {$\checkmark$}  & {}  & {}  & {}    & {Foursquare} \\
 \hline
 {MKM-SR~\cite{meng@SIGIR2020}}    & {SIGIR20'}    &{$\checkmark$}    & {$\checkmark$}   & {}   & {}  & {}   & {}   & {}  & {}  & {$\checkmark$}    & {JingDong} \\  
 \hline
 {MV-RNN~\cite{MV-RNN}}    & {TKDE20'}    &{}    & {}   & {}   & {$\checkmark$}  & {$\checkmark$}   & {}  & {}  & {}  & {}    & {Private} \\  
 \hline
 {ESRM-KG~\cite{liu@WWW2020}}    & {WWW20'}    &{}    & {}   & {}   & {$\checkmark$}  & {}    & {}   & {} & {}  & {}    & {Private} \\
  \hline
  
  {NOVA~\cite{liu@AAAI2021}}    & {AAAI21'}    &{$\checkmark$}    & {}   & {}   & {}  & {}   & {}  & {$\checkmark$}  & {}  & {}    & {MovieLens} \\  
 \hline
  {CBML~\cite{song@CIKM2021}}    & {CIKM21'}    &{$\checkmark$}    & {$\checkmark$}   & {}   & {}  & {}   & {}  & {}  & {}  & {}    & {Yoochoose, Diginetica} \\  
 \hline
 {CoCoRec~\cite{cai@SIGIR2021}}    & {SIGIR21'}    &{$\checkmark$}    & {}   & {}   & {}  & {}   & {}  & {}  & {}  & {}    & {Beer, Taobao} \\  
 \hline

 {MML~\cite{Pan@CIKM2022}} & {CIKM22'}    &{}    & {}   & {}   & {$\checkmark$}  & {$\checkmark$}   & {}  & {}  & {}  & {}    & {Private} \\  
 \hline
 {NextIP~\cite{Luo@CIKM2022}}    & {CIKM22'}    &{}    & {}   & {}   & {}  & {}   & {}  & {}  & {}  & {$\checkmark$}    & {Tmall} \\   
 \hline
 {EMBSR~\cite{Yuan@ICDE2022}}    & {ICDE22'}    &{}    & {}   & {}   & {}  & {}   & {}  & {}  & {}  & {$\checkmark$}    & {JingDong, Trivago} \\  
 \hline
 {MBHT~\cite{yang@KDD2022}}    & {KDD22'}    &{}    & {}   & {}   & {}  & {}  & {}   & {}  & {}  & {$\checkmark$}    & {Retailrocket} \\   
 \hline
 {UniSRec~\cite{hou@KDD2022}}    & {KDD22'}    &{$\checkmark$}    & {$\checkmark$}   & {}   & {$\checkmark$}  & {}   & {}  & {}  & {}  & {}    & {Amazon} \\  
 \hline
 {CARCA~\cite{Rashed@RecSys2022}}    & {RecSys22'}    &{$\checkmark$}    & {$\checkmark$}   & {}   & {}  & {$\checkmark$}   & {}  & {}  & {}  & {}    & {Amazon} \\  
 \hline
 {GPG4HSR~\cite{Chen@RecSys2022}}    & {RecSys22'}    &{}    & {}   & {}   & {}  & {}   & {}  & {}  & {}  & {$\checkmark$}    & {MovieLens, Yoochoose, Tmall} \\
 \hline
 {M2TRec~\cite{Shalaby@RecSys2022}}    & {RecSys22'}    &{$\checkmark$}    & {}   & {}   & {$\checkmark$}  & {}   & {}   & {}  & {}  & {}    & {Diginetica} \\  
 \hline
 {CoHHN~\cite{CoHHN}}    & {SIGIR22'}    &{$\checkmark$}    & {}   & {$\checkmark$}   & {}  & {}  & {}   & {}  & {}  & {}    & {Amazon, Cosmetics, Diginetica} \\  
 \hline
 {DIF-SR~\cite{xie@SIGIR2022}}    & {SIGIR22'}    &{$\checkmark$}    & {}   & {}   & {}  & {}   & {}  & {}  & {}  & {}    & {Amazon, Yelp} \\  
 \hline
 {MB-STR~\cite{yuan@SIGIR2022}}    & {SIGIR22'}    &{}    & {}   & {}   & {}  & {}   & {}  & {}  & {}  & {$\checkmark$}    & {Taobao, Yelp} \\  
 \hline
 {MGS~\cite{lai@SIGIR2022}}    & {SIGIR22'}    &{$\checkmark$}    & {$\checkmark$}   & {}   & {}  & {}   & {}  & {}  & {}  & {}    & {Diginetica, Tmall} \\  
 \hline
 {DETAIN~\cite{lin@WWW2022}}    & {WWW22'}    &{$\checkmark$}    & {$\checkmark$}   & {$\checkmark$}   & {}  & {}   & {}   & {$\checkmark$}  & {}  & {}    & {Amazon, LastFM} \\  
 \hline

 {MMSR~\cite{Hu@CIKM2023}}    & {CIKM23'}    &{}    & {}   & {}   & {$\checkmark$}  & {$\checkmark$}   & {}  & {}  & {}  & {}    & {Amazon} \\
 \hline
 {M3SRec~\cite{Bian@CIKM2023}}    & {CIKM23'}    &{}    & {}   & {}   & {$\checkmark$}  & {$\checkmark$}    & {}  & {}  & {}  & {}    & {Amazon} \\
\hline
{TASTE~\cite{Liu@CIKM2023Text}}    & {CIKM23'}    &{}    & {}   & {}   & {$\checkmark$}  & {}   & {}  & {}  & {}  & {}    & {Amazon, Yelp} \\
 \hline
 {REKS~\cite{Wu@ICDE2023}}    & {ICDE23'}    &{$\checkmark$}    & {$\checkmark$}   & {}   & {}  & {}  & {}   & {}  & {}  & {}    & {Amazon, MovieLens} \\
 \hline
{CCA~\cite{Huang@ICDM2023}}    & {ICDM23'}    &{}    & {}   & {}   & {}  & {}  & {}  & {}    & {$\checkmark$}  & {}    & {Amazon} \\ 
 \hline

 {Recformer~\cite{Li@KDD2023}}    & {KDD23'}    &{$\checkmark$}    & {$\checkmark$}   & {}   & {$\checkmark$}  & {}   & {}  & {}  & {}  & {}    & {Amazon} \\
 \hline
 {FAPAT~\cite{Liu@NeurIPS2023}}    & {NeurIPS23'}    &{$\checkmark$}    & {$\checkmark$}   & {}   & {}  & {}   & {}  & {}    & {}  & {}    & {Diginetica, Tmall, Private} \\  
 \hline
  {DLFS-Rec~\cite{Liu@RecSys2023}}    & {RecSys23'}    &{$\checkmark$}    & {$\checkmark$}   & {}   & {}  & {}   & {}  & {}  & {}  & {}    & {Amazon} \\
 \hline
 {HPM~\cite{Huang@SIGIR2023}}    & {SIGIR23'}    &{$\checkmark$}    & {}   & {}   & {}  & {}   & {}  & {}  & {}  & {}    & {Amazon} \\  
 \hline
 {KMVG~\cite{Chen@SIGIR2023}}    & {SIGIR23'}    &{$\checkmark$}    & {$\checkmark$}   & {}   & {}  & {}  & {}   & {}  & {}  & {}    & {Amazon, Cosmetics, Yelp} \\
 \hline
 {LP-MRGNN~\cite{WW@TKDE2023}}    & {TKDE23'}    &{}    & {}   & {}   & {}  & {}   & {}  & {}  & {}  & {$\checkmark$}    & {Yoochoose, Taobao, Private} \\  
 \hline
 {MMSBR~\cite{MMSBR}}    & {TKDE23'}    &{}    & {}   & {$\checkmark$}   & {$\checkmark$}  & {$\checkmark$}   & {}  & {}  & {}  & {}    & {Amazon} \\
 \hline
 {BiPNet~\cite{BiPNet}}    & {TOIS23'}    &{$\checkmark$}    & {$\checkmark$}   & {$\checkmark$}   & {}  & {}  & {}   & {}  & {}  & {}    & {Amazon, Cosmetics} \\
\hline 
{DUVRec~\cite{Xue@TOIS2023}}    & {TOIS23'}    &{$\checkmark$}    & {}   & {}   & {}  & {}  & {}   & {}  & {}  & {}    & {Amazon, MovieLens, Steam} \\
\hline 
  
 {DisenPOI~\cite{Qin@WSDM2023}}    & {WSDM23'}    &{}    & {}   & {}   & {}  & {}   & {$\checkmark$}  & {}  & {}  & {}    & {Foursquare} \\
 \hline
 {MMMLP~\cite{MMMLP}}    & {WWW23'}    &{}    & {}   & {}   & {$\checkmark$}  & {$\checkmark$}   & {}  & {}  & {}  & {}    & {MovieLens} \\  
 \hline
 
 % {KGDPL~\cite{Liu@TKDE2024}}    & {TKDE}    &{}    & {}   & {}   & {$\checkmark$}  & {$\checkmark$}   & {}  & {}  & {}  & {}    & {Amazon,LastFM,MovieLens,Yelp} \\
  % \hline
  
 {DWSRec~\cite{Zhang@AAAI2024}}    & {AAAI24'}    &{}    & {}   & {}   & {$\checkmark$}  & {}   & {}  & {}  & {}  & {}    & {Amazon} \\
 \hline
 {DIMO~\cite{DIMO}}    & {SIGIR24'}    &{}    & {}   & {}   & {$\checkmark$}  & {$\checkmark$}  & {}   & {}  & {}  & {}    & {Amazon, Instacart} \\
 \hline
 {LLM4ISR~\cite{Sun@SIGIR2024}}    & {SIGIR24'}    &{}    & {}   & {}   & {$\checkmark$}  & {}   & {}  & {}  & {}  & {}    & {Amazon,MovieLens} \\
 \hline
 {GHTID~\cite{Li@WSDM2024}}    & {WSDM24'}    &{}    & {}   & {}   & {}  & {}   & {}  & {}  & {}  & {$\checkmark$}    & {MovieLens, Yoochoose, Tmall} \\  
 \hline
 {MSSR~\cite{Lin@WSDM2024}}    & {WSDM24'}    &{$\checkmark$}    & {}   & {}   & {}  & {}   & {}  & {}  & {}  & {}    & {Amazon, Yelp} \\  
 \hline
 {ASIF~\cite{WangS@WWW2024}}    & {WWW24'}    &{$\checkmark$}    & {$\checkmark$}   & {}   & {}  & {}   & {}  & {}  & {}  & {}    & {Amazon,Yelp,Private} \\  
 \hline
 {END4Rec~\cite{Han@WWW2024}}    & {WWW24'}    &{}    & {}   & {}   & {}  & {}   & {}  & {}  & {}  & {$\checkmark$}    & {Diginetica, Yoochoose, Taobao} \\  
 \hline
 {LLM-TRSR~\cite{Zheng@WWW2024}}    & {WWW24'}    &{}    & {}   & {}   & {$\checkmark$}  & {}   & {}  & {}  & {}  & {}    & {Amazon} \\
 \hline
 {M-KGHT~\cite{Wang@WWW2024}}    & {WWW24'}    &{$\checkmark$}    & {$\checkmark$}   & {$\checkmark$}   & {$\checkmark$}  & {}   & {}  & {}  & {}  & {}    & {Diginetica, Yoochoose, Private} \\
 \hline
 {SLIM~\cite{Wang@WWW2024LLM}}    & {WWW24'}    &{}    & {}   & {}   & {$\checkmark$}  & {}   & {}  & {}  & {}  & {}    & {Amazon} \\
 \hline
 {SAID~\cite{Hu@WWW2024}}    & {WWW24'}    &{}    & {}   & {}   & {$\checkmark$}  & {}   & {}  & {}  & {}  & {}    & {Amazon} \\

% \midrule
% \multicolumn{11}{p{\linewidth}}{Amazon contains many sub-datasets covering various domains such as Beauty, Cellphones and Accessories, Toys and Games and so on. Similarly, JingDong also contains two sub-datasets Applicances and Computers}          \\ 
\bottomrule
\end{tabular}
% \vspace{-0.1in}
% \caption{An overview of the side information-driven session-based recommendation approaches including the model names, the information types they used and corresponding datasets.}
% \vspace{-0.1in}
\label{approach}
\end{table*}

\subsection{Time-driven SBR}\label{timesbr}
{Time} has been recognized as a crucial indicator of users' changing preferences and has been widely studied~\cite{GRU4Rec,Alexander@arXiv2017,NARM}. There are two primary manners to integrate time information into SBR: \textit{temporal order} and \textit{time interval}. 

Temporal order models only take the order of user-item interactions into account, emphasizing the relative position of interactions within a session. Specifically, these methods can be further divided into two categories: implicit and explicit modeling. On one hand, implicit methods rely on RNN and its variants, which inherently process data in sequential order, to implicitly capture sequential patterns among items without needing extra time-related inputs, such GRU4Rec~\cite{GRU4Rec}, NARM~\cite{NARM} and RepeatNet~\cite{RepeatNet}. On the other hand, explicit efforts directly inject position embeddings into original item embeddings to achieve time-aware item embeddings including~\cite{Garg@SIGIR2019} with KNN, ~\cite{LESSR,FGNN,Li@CIKM2022} with GNN and~\cite{SASRec,STAMP} with self-attention. In contrast, time interval models focus on the specific duration between interactions, assuming that longer browsing times could indicate stronger user preferences.
These methods usually discretize continuous time intervals into one-hot encoding and leverage self-attention~\cite{Li@WSDM2020, Wu@WWW2020, Chen@CIKM2022} or contrastive learning~\cite{Tian@CIKM2022,Dang@AAAI2023} to capture temporal patterns. 

Despite the extensive investigation of time information, most methods primarily concentrate on straightforward sequential dependencies, typically from the single perspective of item order. In reality, time information offers rich insights into user behavior from various perspectives. For example, considering time from a daily perspective can reveal users' habits, such as a tendency for fast-food at noon and full- meals at night. Incorporating diverse time views, such as daily routines, seasonal changes, or even yearly trends, could unlock deeper insights into user preferences and lead to more contextually appropriate recommendations.

% Despite extensive studies, time information is often incorporated separately. We believe that a joint modeling of time with other types of information could unveil deeper insights into user actions. For instance, integrating time information with item colors from images could reveal seasonal preferences, such as a tendency for lighter-colored clothing in summer and darker shades in winter.
% Such an integration has the potential to significantly enhance the performance of SBR models.

\subsection{Category-driven SBR}\label{catesbr}

The category feature suggests a user’s hierarchical decision-making process, offering the potential to narrow down the range of candidate items. As the pioneering efforts incorporate item categories into this task, CoCoRec~\cite{cai@SIGIR2021} relies on self-attention layers to explore user in-category preferences for improving recommendation performance. Other approaches model item ID sequences and category sequences simultaneously using GNN~\cite{lai@SIGIR2022} or self-attention~\cite{Zhou@CIKM2020} for comprehensive session representation learning. 

Nevertheless, existing methods typically convert item categories into one-hot encoding to serve as inputs for neural networks. This paradigm, while straightforward, falls short of capturing the hierarchical associations among categories, where this hierarchical feature is important in refining user intents. As a result, it leads to a loss of valuable contextual information and impedes the effective utilization of category information in the task.

\subsection{Brand-driven SBR}\label{brandsbr}
The brand serves as a distinctive indicator of a manufacturer's reputation among available side information. It holds significant influence over user behavior, particularly in cases driven by the `fandom economy' as highlighted in economic research. Despite its potential impact, unfortunately, none of the existing methods have specifically focused on incorporating brand information into the task of \baby. Instead, brand data is usually bundled with other types of side information within current methods. A common practice involves combining brand attributes with categories, often employing separate self-attention layers to model item ID and side information independently, thereby enhancing the representation of sessions and improving recommendation performance~\cite{zhang@IJCAI2019,song@CIKM2021,Liu@RecSys2023}.   

\subsection{Price-driven SBR}\label{pricesbr}
Price, as a pivotal factor users concern when purchasing items, plays a crucial role in influencing their decisions. Surprisingly, limited research has incorporated price information into the domain of SBR. A notable attempt is made by CoHHN~\cite{CoHHN}, which constructs a heterogeneous hypergraph encoding item ID, price, and categories simultaneously to simultaneously capture a user's price and interest preferences. Building on this, BiPNet~\cite{BiPNet} introduces additional item brand information to further enhance the model’s ability to capture user sensitivity to price variations. Given its unique and critical role in the user decision-making process, we believe that there is considerable potential for further research to more effectively integrate price factors into solutions of \baby.

\subsection{Text-driven SBR}\label{textsbr}
As a primary way to display items on the website, text elements, including item titles and descriptions, enable users to quickly grasp the essence of items, directly influencing their choices. Various approaches have been developed to incorporate this textual data into SBR, aiming to leverage its rich semantics for pinpointing user intents. For instance, ESRM-KG~\cite{liu@WWW2020} enhances recommendation performance through a keyword generation task which is designed to uncover shared semantics among items with distinct IDs. Some recent models, like UniSRec~\cite{hou@KDD2022} and Reformer~\cite{Li@KDD2023}, introduce pre-trained NLP techniques into user behavior modeling, aiming at transferring linguistic knowledge to recommendation domain. These approaches typically represent an item via modeling its text, train models following the training-fine-tuning paradigm, and achieve item and sequence representation learning. Due to its remarkable success in the field of NLP, Transformer architecture is prevalent in these text-driven methods, as seen in models such as M2TRec~\cite{Shalaby@RecSys2022} and DWSRec~\cite{Zhang@AAAI2024}. Additionally, increasing efforts attempt to fine-tune Large Language Models based on item sequences for LLM's adaption on recommendation task. In these solutions, an item is also represented by its text. In this emerging paradigm for recommendation task, recent methods usually focus on constructing effective prompts based on item text to unleash LLM's power on capturing user intents. Representative instances include summarization prompts in LLM-TRSR~\cite{Zheng@WWW2024}, chain-of-thought prompting in SLIM~\cite{Wang@WWW2024LLM}, as well as generative prompts in LLM4ISR~\cite{Sun@SIGIR2024}.

Despite the advancements in text-driven SBR methods, existing methods, no matter based on neural networks or LLMs, typically process item text data as a whole, \ie they often convert lengthy descriptions into a single dense embedding. This paradigm overlooks to distinguish between relevant and irrelevant information and fails to prioritize key features that are critical to user preferences. As a result, valuable semantic details that could offer insights into user intents are lost, thereby limiting the model's ability to generate effective recommendations.

% \noindent\textbf{Description} 

\subsection{Image-involved Multi-modal SBR}\label{imagesbr}
Images offer a vivid portrayal of an item's appearance, highlighting visual features like color and style. Due to their intuitive and appealing nature, images typically occupy a prominent position on websites. In practice, images are frequently combined with text, formulating multi-modal recommendation tasks. A representative approach specially designed for multi-modal SBR is MMSBR~\cite{MMSBR}. This model accounts for distinct user behavior patterns across different modalities, aiming to capture user preferences holistically. In addition, various techniques have been utilized to fuse multi-modal embeddings for a comprehensive understanding of user intents, including simple weighted sum~\cite{P-RNN}, Multi-Layer Perceptron (MLP)~\cite{MMMLP}, self-attention mechanism~\cite{Pan@CIKM2022,Bian@CIKM2023} and GNN~\cite{Hu@CIKM2023}. 

Despite the promising potential, multi-modal SBR is still in its infancy. Several critical challenges remain in this rapidly evolving domain, such as handling modality heterogeneity, addressing noisy information inherent in modalities, uncovering user preferences unique to each modality, and generating explainable recommendations.

\subsection{Point-of-interest SBR}
Address information, which provides the geographical context of items, is essential for a specialized recommendation scenario known as Point-of-Interest (POI) SBR~\cite{Zhang@IS2020}. In the task of POI SBR, user actions are influenced by both sequential behaviors and geographical proximity. Recent research has increasingly focused on integrating spatial and temporal information to enhance Point-of-interest SBR. For instance, GeoSAN~\cite{Lian@KDD2020} emphasizes the exact GPS coordinates of locations, while STiSAN~\cite{Wang@ICDE2022} prioritizes time intervals between consecutive visits. DisenPOI~\cite{Qin@WSDM2023} takes a step further by disentangling sequential and geographical relationships using contrastive learning techniques, aiming to enhance the quality of embeddings for more accurate recommendations.

\subsection{Rating-driven SBR}\label{ratingsbr}
Rating data provides a direct and explicit measure of user preferences towards items, offering valuable insights into the user attitudes. While rating prediction was a primary focus in the early stages of recommender system's research, its application has been somewhat overlooked in the realm of session-based recommendation. There are only a few methods taking rating data into account in the topic. For instance, NOVA~\cite{liu@AAAI2021} presents a non-invasive fusion mechanism based on self-attention to fuse ID, rating, and category information for user sequential behavior modeling. In this setup, the fused information serves as the Key and Query vectors, while item IDs act as the Value vectors. Given that ratings can explicitly deliver user preferences, we argue that this information warrants deeper exploration in SBR. For example, replacing implicit user actions with explicit ratings as supervised signals could provide a more direct and meaningful guide for model training. This approach would better leverage user-stated preferences, providing clearer indications of user intent and enhancing the effectiveness of SBR models.

\subsection{Review-driven SBR}\label{reviewsbr}
Reviews, posted by users after consuming products or services, offer detailed insights into both user preferences and item characteristics~\cite{FineRec}. To this end, incorporating user-item reviews into recommendation models has attracted increasing attention recently. As the pioneering work in this topic, RNS~\cite{RNS} represents items with the reviews they received, employs an attention mechanism to learn a user's short- and long-term interests, and merges these interests to strengthen model’s expressive ability. DER~\cite{Chen@AAAI2019} focuses on profiling items by extracting informative reviews while also considering time intervals between actions to better capture dynamic user preferences, improving both accuracy and explainability of its suggestions. CCA~\cite{Huang@ICDM2023} further exploits the complementary nature of individual reviews (\eg a single review of an item) and aggregated reviews (\eg all reviews an item received) to maximize the rich information of review data. 

A major challenge in review-driven recommendation lies in handling review heterogeneity. Specifically, distinct users may provide significantly different, even contradictory, opinions on the same item, making it difficult to learn a consistent and robust representation for that item. Besides, existing approaches often treat reviews as a singular unit, aiming to capture user preferences from the review as a whole. However, within a single review, a user may express conflicting opinions over different attributes of an item. For instance, a review for a T-shirt might say: `The color is perfect, while I just do not like the patterns'. This highlights the need for future research to develop methods that can analyze user-item reviews in a more fine-grained, attribute-specific manner. 

% The rich semantics contained in reviews hold immense potential for improving SBR. For example, the specific preferences revealed in user reviews could be leveraged to provide explanations alongside recommendations, contributing to transparent and trustworthy recommendations. 

\subsection{Multi-behavior SBR}\label{behaviorsbr}
Reflecting diverse user intents, different behavior types have become a focal point in SBR recently, leading to the emergence of a new recommendation task known as `Micro-Behavior' or `Multi-behavior' session-based recommendation. Current solutions in this emerging area generally involve constructing separate sequences for items and behavior types, such as item sequences and operation sequences. These sequences are then jointly analyzed to forecast the user's next actions, accounting for the unique context and purpose associated with each type of behavior. For instance, MKM-SR~\cite{meng@SIGIR2020} applies GNN and GRU to handle item sequences and operation sequences respectively, while EMBSR~\cite{Yuan@ICDE2022} processes item sequences with GNN and uses self-attention for operation sequences. Recent advances have also sought to mitigate the noise that arises from the incorporation of various behaviors, as seen in END4Rec~\cite{Han@WWW2024} and GHTID~\cite{Li@WSDM2024}.

At its initial stage, multi-behavior SBR suffers from several limitations. Firstly, by modeling item and behavior sequences separately, current methods are limited to capturing only the intra-dependencies within a single type of sequence. This approach overlooks the complex inter-dependencies that exist between different types of sequences, which are crucial for a comprehensive understanding of user behaviors. Secondly, the choice of neural architectures such as RNNs, GNNs, or Transformers significantly influences recommendation performance. However, current models typically select these structures based on empirical guidelines, making it challenging to identify the optimal architecture for a given scenario. Thirdly, current methods often focus on a predefined set of behavior types, which restricts their adaptability to new or evolving behaviors. This rigidity can impair the model's performance in real-world applications, where user interactions are continually evolving, requiring a more flexible and robust approach to handle the changing behavioral landscape.

% Han@WWW2024

% Although achieving impressive performance, these methods focus solely on behavior types, while suffering from the lack of integration of other available information. For example, by observing that a user frequently clicks on flat shoes but purchases high heels, a more insightful understanding of her fine-grained interests can be achieved. Such integrated approach could lead to more accurate and personalized SBR.

\subsection{Others}\label{othersbr}
In addition to the aforementioned side information, recent studies in SBR have begun to delve into more complex types of data. This includes not just item side information, but also the relations between different types of information. For instance, a `belong-to' relation might indicate that an item belongs to a particular category or brand. Knowledge Graphs (KGs), consisting of entity nodes and relation edges, are especially well-suited for representing such intricate relationships. In this context, an entity in a KG could be an item, brand, or category, while relations represent connections like `belong-to' between these entities. 
Current methods, such as KSR~\cite{Huang@SIGIR2018},  REKS~\cite{Wu@ICDE2023}, KMVG~\cite{Chen@SIGIR2023} and M-KGHT~\cite{Wang@WWW2024}, integrate KGs into SBR to leverage the wealth of knowledge embedded in these graphs, aiming at generating more informative item embeddings for better recommendations.

A key challenge in KG-based SBR lies in the requirement for a well-structured KG, which often demands extensive domain expertise and considerable manual effort to construct and maintain. This issue seriously limits the development of KG in SBR.

% Overall, in our summary of the current research progress about side information-driven SBR, we have categorized the studies based on the types of information they utilize. Our intention is that this taxonomy will provide a convenient reference for future research in locating relevant studies, replicating existing methods, conducting comparative analyses, and identifying opportunities for further enhancements.

\section{Open Challenges and Future Directions}\label{openf}
In recent years, side information-driven session-based recommendation has witnessed significant growth, driven by advancements in neural network technologies ranging from RNN, CNN, attention mechanism, GNN and contrastive learning to LLMs. After offering a detailed review of the research practices in \baby, we also identify and discuss several open research directions, highlighting opportunities for future breakthroughs and advancements in the field.
% In this section, we identify and discuss several open research directions, setting the stage for future breakthroughs and advancements in the field.

\subsection{Strengthen the Utilization of Available Information in SBR}
As previously discussed, each type of information, ranging from time and price to images and reviews, plays a distinct role in capturing user preferences. These diverse data sources are instrumental in uncovering specific user interests and delivering personalized services.  Unfortunately, as noted in the section of Research Progress (Section~\ref{ResearchP}), certain types of information, such as brand, price, and reviews, remain relatively under-explored in current SBR research. To address this gap, more efforts should be dedicated to incorporating these less-exploited data types into SBR, leveraging their unique value to enhance the understanding of user behaviors in this challenging task.

\subsection{Joint Incorporation of Various Information into SBR} 
Current methods in SBR, as outlined in Table~\ref{approach}, predominantly utilize only a limited subset of the available side information. The potential of combining multiple diverse information sources remains largely unexplored. This integration of various information could significantly enhance session data and provide a deeper insight into user behaviors. For example, merging time information with color attributes derived from item images could uncover seasonal trends in user interests, such as a tendency towards lighter colors in summer apparel and darker tones in winter wear. Obviously, such joint incorporation endows recommender systems with the potential to capture user preferences in a fine-grained manner, enabling the delivery of highly personalized and, at times, unexpectedly relevant suggestions. Furthermore, leveraging a broader range of information sources can effectively mitigate the data sparsity challenge that is inherent in SBR.

\subsection{Side Information Enhanced Cold-start SBR}
The cold-start problem remains a persistent challenge in recommender systems, where the systems fall short of offering suggestions about new items. Despite its prominence, this long-standing issue has received limited attention within SBR. The root cause of the cold-start problem lies in the lack of collaborative relationships between new and existing items, \ie item-level co-occurrence patterns. Consequently, the mainstream ID-based methods struggle to recommend newly added items effectively. Fortunately, the incorporation of side information presents a promising avenue to tackle this issue by enabling recommendations based on collaborative relations at fine-grained feature-level instead of coarse-grained item-level. For instance, it becomes feasible to recommend a new action movie starring Jackie Chan to a user who has previously enjoyed films featuring him, by leveraging detailed side information like the cast. This approach allows the system to draw meaningful connections and deliver relevant suggestions, even in the absence of direct historical data.

\subsection{Side Information Driven Explainable SBR}
Providing explanations alongside recommendations can significantly enhance user satisfaction and trust in recommender systems. Nevertheless, explainable SBR has not yet received extensive attention. Side information, with its rich semantics, serves as a valuable source for uncovering user specific interests, contributing to generating personalized and insightful explanations. For instance, as illustrated in the second row of ~\fig~\ref{intro}, a user's preference for Marvel-related items can be inferred from the images and text of his previously purchased items. Based on these fine-grained interests, the system can generate personalized explanations such as, `You have purchased items featuring Marvel themes, so we recommend a T-shirt with a Marvel logo'. This approach offers clear justifications for its suggestions, enhancing system transparency and user engagement. As a result, leveraging side information to create explainable SBR would emerge as an important direction for future research.

% \subsection{Side information enhanced cross-field SBR}
% Cross-field recommendation seeks to transfer knowledge from a source domain, rich in user-item interactions, to a target domain where such interactions are sparse, enhancing the user experiences in the target field. The item side information is ideally suited for the transferring operations since user fine-grained preferences about such information are relatively stable and independent from user-item interaction data. However, few efforts have been dedicated to the direction.

\subsection{LLM-based SBR with Side Information}
Large Language Models (LLMs) have shown remarkable capabilities across various tasks and are increasingly being explored in the field of recommender systems. However, their initial applications in this domain have not been as successful as anticipated. We contend that while LLMs encapsulate rich general knowledge, they struggle to learn the collaborative signals reflected by item IDs, which are of great significance under the scenario of recommendation. This challenge is particularly evident in SBR, where sessions often involve limited user actions. As a result, current LLM-based methods have yet to match the performance of mainstream ID-based neural models. Fortunately, side information discussed in this survey, especially item text, presents a significant opportunity to unleash the full potential of LLMs in SBR. Since LLMs are adept at handling modality-rich data, like item titles and images, they can better grasp user preferences hidden in such information. Therefore, further investigation into LLM-based SBR with side information is not only warranted but essential for advancing this field.

\subsection{Side Information-based Benchmarks for SBR}
% As emphasized in previous sections, datasets are a vital resource in the domain of \baby. However, 
As outlined in Table~\ref{datasets}, none of the existing datasets cover all types of information required for a holistic evaluation of methods in this field. This limitation poses a significant challenge to conducting fair performance comparisons. For instance, a method may excel in leveraging behavior types using `Behavior' information on the Cosmetics, while another one might concentrate on capturing user preferences from item images using `Images' information on the Amazon. Lacking a dataset including both `Behavior' and `Images' information simultaneously, it becomes impossible to directly compare the performance of these methods. This gap underscores the urgent need for high-quality datasets that integrate diverse side information. However, a major challenge in collecting such data arises from privacy policies enforced by various online platforms. We encourage researchers and platform providers to consider publicly releasing such datasets, thereby supporting the development of this promising research topic.

\section{Conclusions}\label{conclusion}
The session-based recommendation is a hot-spot research topic due to its substantial application value in offering personalized services for anonymous users based on their limited actions. In recent times, there has been a surge in efforts to integrate side information into SBR as a means to address its inherent issues of data sparsity, giving rise to a promising topic of side information-driven session-based recommendation. In this domain, data plays a crucial role as it directly influences the design, implementation, and evaluation of recommendation algorithms. As a result, we elaborated on side information-driven SBR from a data-centric perspective in this work. This survey commences with formulating the paradigm of \baby, details the available benchmarks, discusses the data characteristics and utility, illustrates the usage of various side information, comprehensively reviews the research progress, and points out future directions. We sincerely hope that our insights and analyses will serve as a valuable resource for researchers and practitioners, encouraging further innovations and developments in this vibrant topic.

% \section*{Acknowledgments}

\newpage

%% The file named.bst is a bibliography style file for BibTeX 0.99c

\bibliographystyle{IEEEtran}
% argument is your BibTeX string definitions and bibliography database(s)
%\bibliography{IEEEabrv,../bib/paper}
%
% <OR> manually copy in the resultant .bbl file
% set second argument of \begin to the number of references
% (used to reserve space for the reference number labels box)

\bibliography{ref}

% \begin{IEEEbiographynophoto}{Jane Doe}
% Biography text here without a photo.
% \end{IEEEbiographynophoto}
\vspace{-40px}
\begin{IEEEbiography}[{\includegraphics[width=1in,height=1.25in,clip,keepaspectratio]{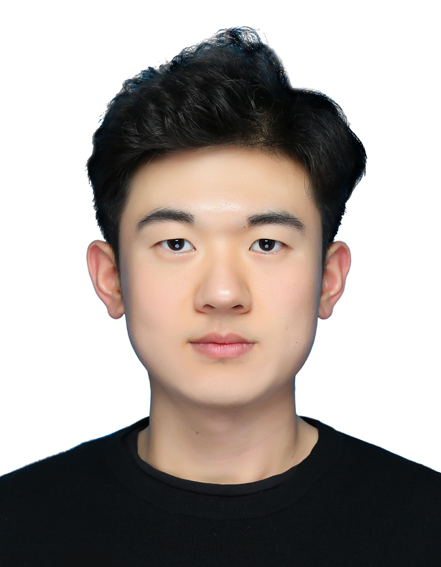}}]{Xiaokun Zhang}
is currently a postdoc research fellow in the Department of Computer Science, City University of Hong Kong. He received his B.S., M.S. and Phd degrees in the School of Computer Science and Technology, Dalian University of Technology. His research interests include data mining and natural language processing, especially focusing on intelligent recommender systems.
\end{IEEEbiography}
\vspace{-40px}
\begin{IEEEbiography}[{\includegraphics[width=1in,height=1.25in,clip,keepaspectratio]{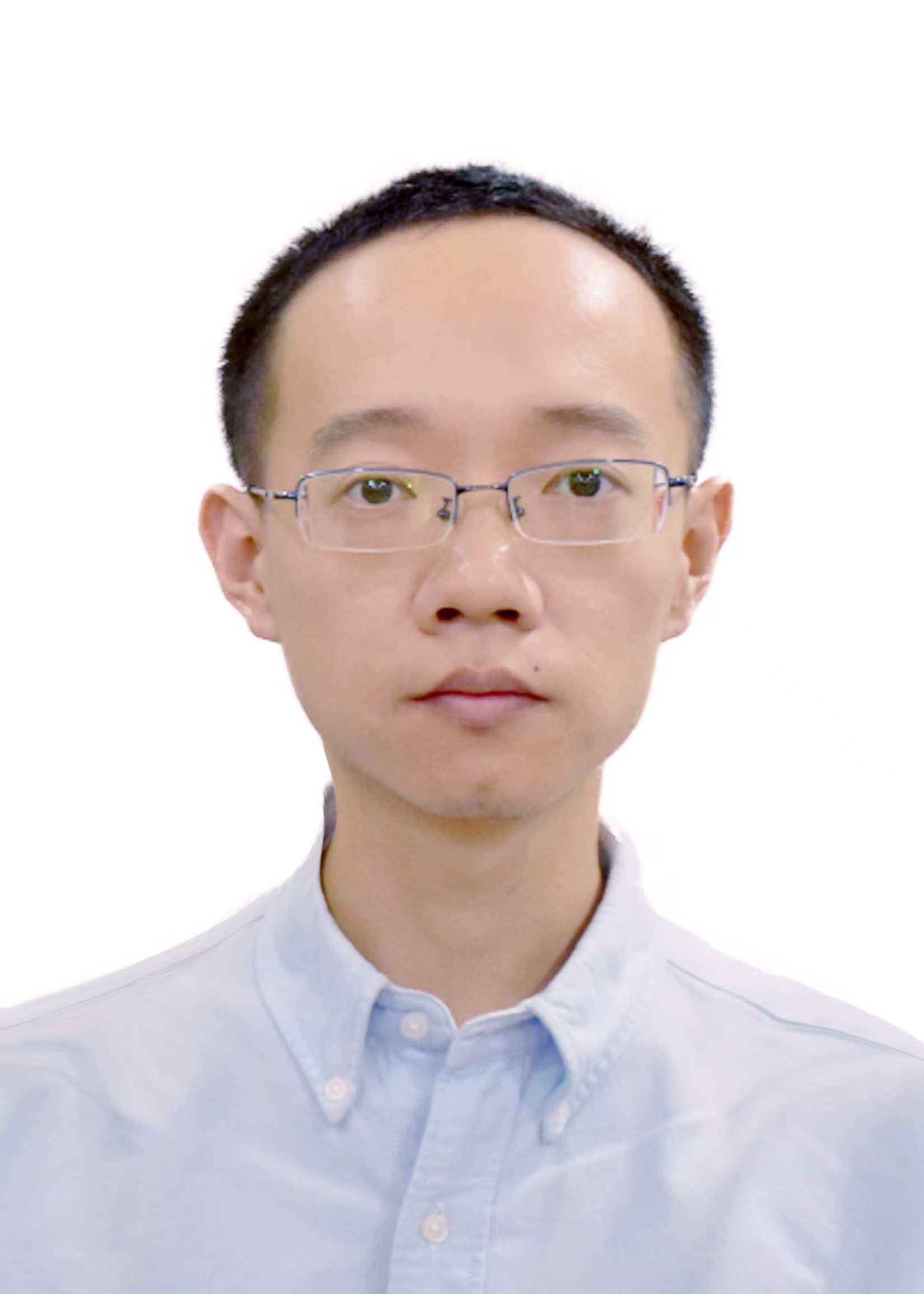}}]{Bo Xu}
received the BSc and PhD degrees from the Dalian University of Technology, China, in 2011 and 2018. He is currently an associate professor in School of Computer Science and Technology of Dalian University of Technology. His current research interests include information retrieval and natural language processing.
\end{IEEEbiography}
\vspace{-40px}
\begin{IEEEbiography}[{\includegraphics[width=1in,height=1.25in,clip,keepaspectratio]{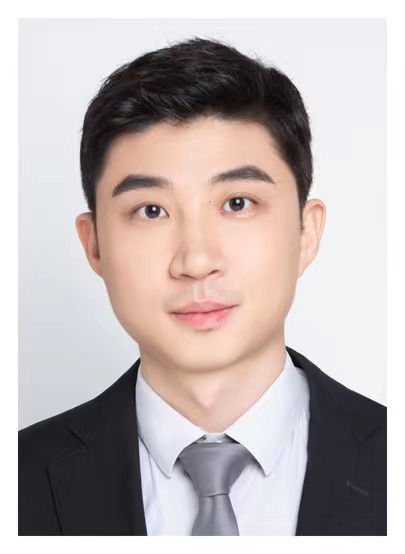}}]{Chenliang Li}
is a full Professor with School of Cyber Science and Engineering, Wuhan University China. His research areas include Information Retrieval, Recommendation
System, Natural Language Processing and Social Computing. He has published over 80 papers in leading conferences and journals, and was the recipient of SIGIR 2016 Best Student Paper Award Honorable Mention. Currently, He serves as an Associate Editor for ACM TOIS, ACM TALLIP, and an editorial board member of JASIST and IPM. He has been a PC member of many leading conference, such as SIGIR, WWW, ACL, WSDM, AAAI, CIKM.
\end{IEEEbiography}
\vspace{-40px}
\begin{IEEEbiography}[{\includegraphics[width=1in,height=1.25in,clip,keepaspectratio]{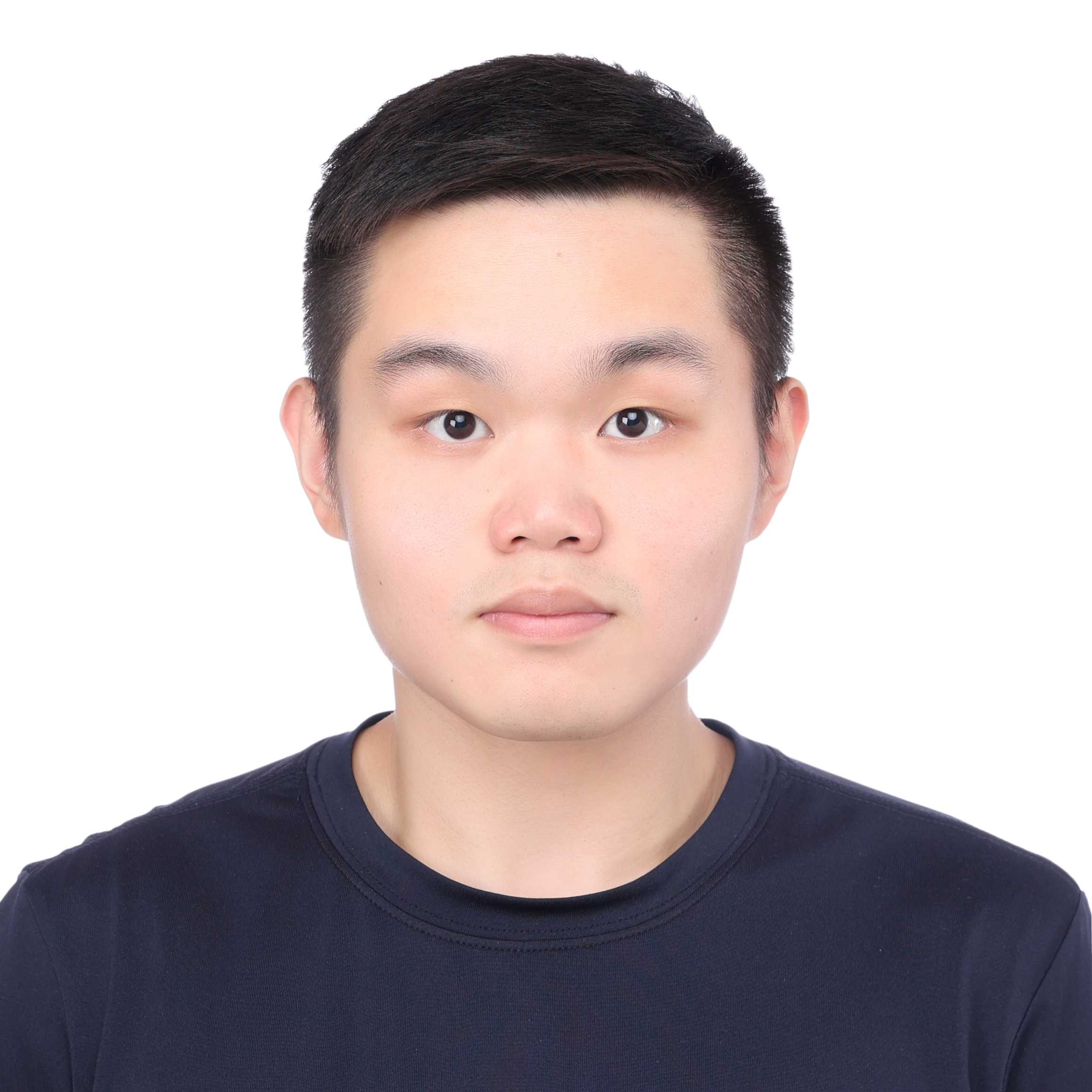}}]{Bowei He} 
is a Ph.D. candidate in Computer Science at City University of Hong Kong. His research focuses on advanced ML (e.g., RL, continual learning, and Auto-ML) and causal inference for data mining and information retrieval. His recent interests include efficient and personalized large language model. He has published papers on KDD, WWW, NeurIPS, ICDE, TKDE, CVPR, ICCV, CIKM, ICDM, UAI, SDM, etc. He interned and collaborated with Didi AI Labs, Huawei Noah's Ark Lab, Alibaba, Kuaishou, and Tencent Financial Technology.
\end{IEEEbiography}
\vspace{-40px}
\begin{IEEEbiography}[{\includegraphics[width=1in,height=1.25in,clip,keepaspectratio]{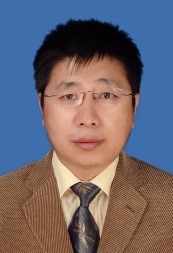}}]{Hongfei Lin}
received the BSc degree from the Northeastern Normal University in 1983, the MSc degree from the Dalian University of Technology in 1992, and the PhD degree from the Northeastern University in 2000. He is currently a professor in School of Computer Science and Technology at the Dalian University of Technology. He has published more than 500 research papers in various journals, conferences, and books. His research interests include information retrieval, text mining for biomedical literatures, biomedical hypothesis generation, information extraction from huge biomedical resources, learning-to-rank. He is the director of Information Retrieval Lab. at Dalian University of Technology.
\end{IEEEbiography}
\vspace{-40px}
\begin{IEEEbiography}[{\includegraphics[width=1.0\columnwidth]{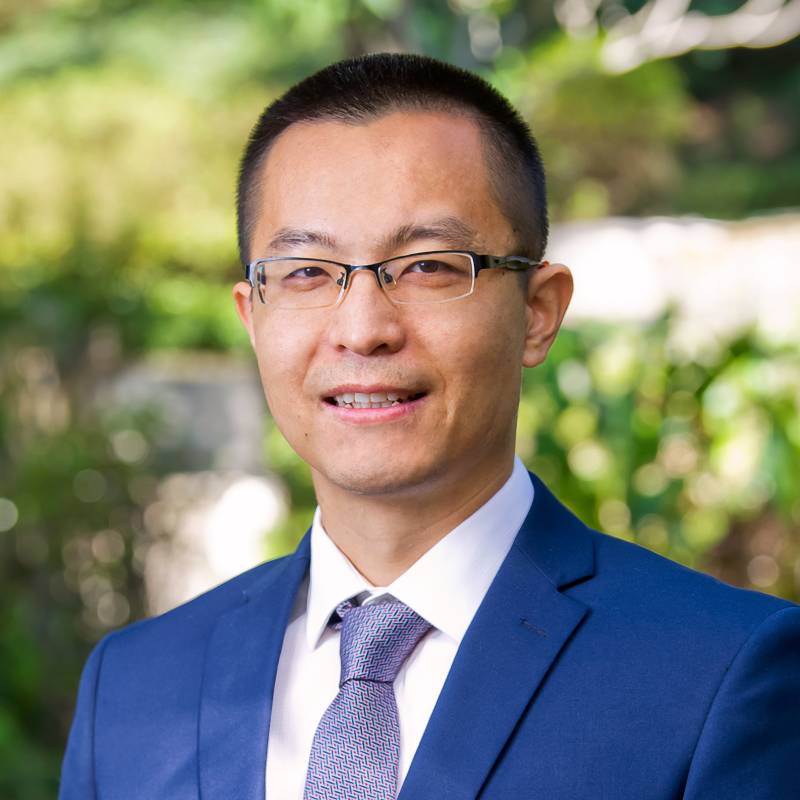}}]{Chen Ma} is currently an Assistant Professor in the Department of Computer Science, City University of Hong Kong. His research interests lie in the theory of data mining and machine learning and their applications in recommender systems, knowledge graphs, social computing, and social good. He received his B.S. and M.S. degrees in Software Engineering from Beijing Institute of Technology in 2013 and 2015, respectively. He received his PhD degree in Computer Science from McGill University.
\end{IEEEbiography}
\vspace{-40px}
\begin{IEEEbiography}[{\includegraphics[width=1in,height=1.25in,clip,keepaspectratio]{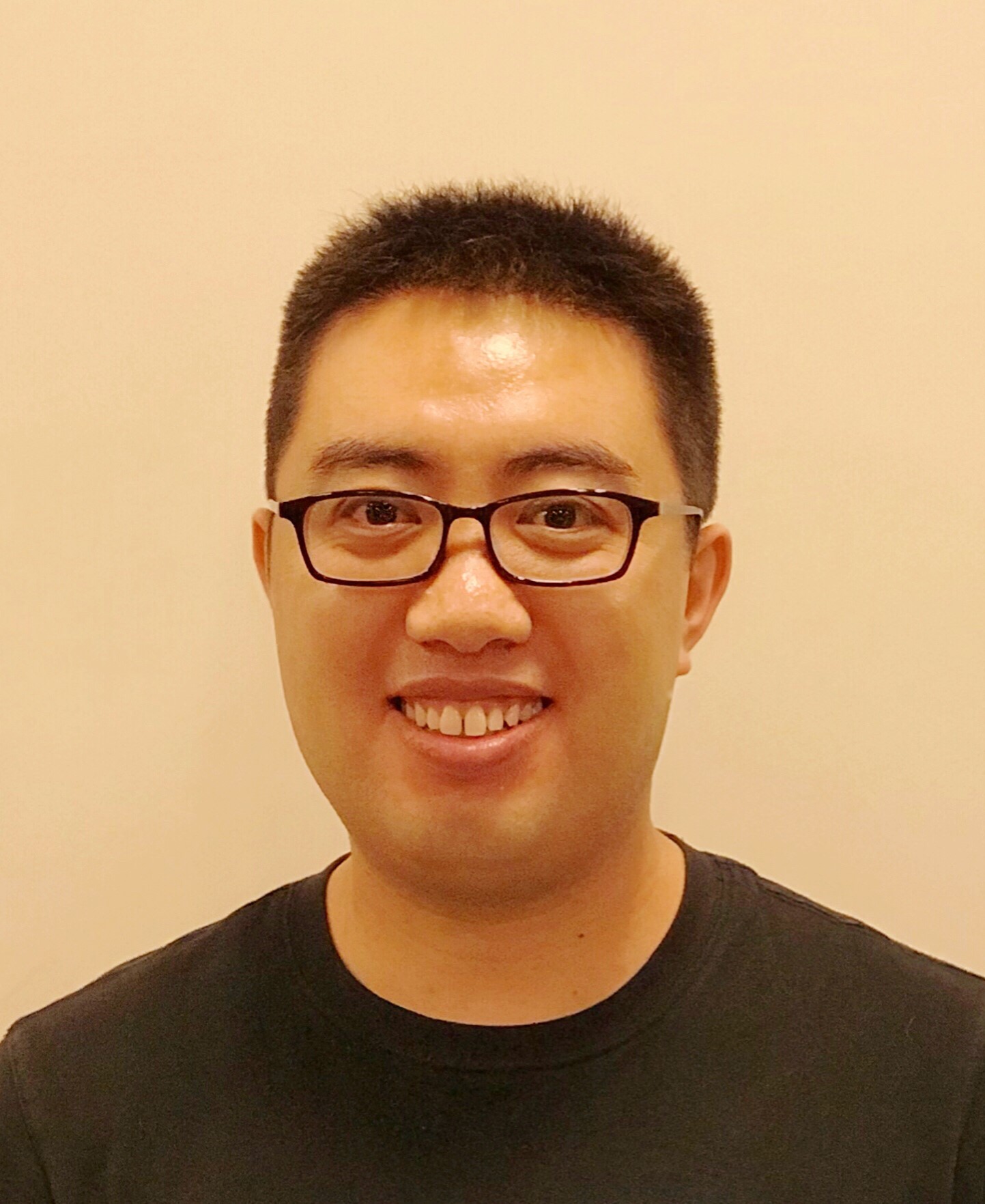}}]{Fenglong Ma}
is an assistant professor in the College of Information Sciences and Technology at the Pennsylvania State University. He received his Ph.D. from the Department of Computer Science and Engineering, University at Buffalo (UB) in 2019. His research interests lie in data mining and machine learning, with an emphasis on mining health-related data. His research interests include natural language processing, federated learning, and AI security.
\end{IEEEbiography}

\end{document}